\documentclass[preprint2]{aastex}

\slugcomment{submitted to Astrophysical Journal}

\shorttitle{Spectro-photometry of galaxies in the Virgo cluster}
\shortauthors{Gavazzi et al.}

\begin{document}


\title{Spectro-photometry of galaxies in the Virgo cluster.
I: The star formation history\footnote{Based on observations collected  
at the  Observatoire de Haute Provence (OHP) (France), 
operated by the CNRS, France, and at the
European Southern Observatory (Chile) (programme 66.B-0026)}}


\author{G. Gavazzi, C. Bonfanti, G. Sanvito\altaffilmark{}}
\affil{Universit\`a degli Studi di Milano - Bicocca, P.zza delle
Scienze 3, 20126 Milano, Italy}
\author{A. Boselli\altaffilmark{}}
\affil{Laboratoire d'Astrophysique de Marseille, Traverse du Siphon, 
F-13376, Marseille Cedex 12, France}
\and
\author{M. Scodeggio\altaffilmark{}}
\affil{Istituto di Fisica Cosmica ``G. Occhialini'', CNR, 
via Bassini 15, 20133, Milano, Italy}




\begin{abstract}

As a result of an extensive observational campaign targeting the Virgo
cluster, we obtained integrated (drift-scan mode) optical spectra and
multiwavelength (UV,U,B,V,H) photometry for 124 and 330 galaxies,
respectively, spanning the whole Hubble sequence, and with $m_p \leq
16~(M_p \leq -15)$.  These data were combined to obtain galaxy
Spectral Energy Distributions (SEDs) extending from 2000 to 22000 \AA.
By fitting these SEDs with synthetic ones derived using Bruzual \&
Charlot population synthesis models we try to constrain
observationally the Star Formation History (SFH) of galaxies in the
rich cluster of galaxies nearest to us.  Assuming a Salpeter IMF and
an analytical form for the SFH, the fit free parameters are: the age
($T$) of the star formation event, its characteristic time-scale
($\tau$) and the initial metallicity ($Z$).
In this work we test the (simplistic) case where all galaxies 
have a common age $T$=13 Gyr, exploring a SFH with "delayed" 
exponential form (which we call "a la Sandage"), thus allowing 
for an increasing SFR with time.  
This SFH is consistent with the full range of observed 
SEDs, provided that the characteristic time-scale $\tau$ is let 
free to vary between 0.1 (quasi instantaneous burst) and 25 Gyr 
(increasing SFR) and $Z$ between 1/50 and 2.5 Z$\odot$.
Elliptical galaxies (including dEs) are best fitted with short time
scales ($\tau \sim 3$ Gyr) and metallicity varying between 1/5 and 1
Z$\odot$. The model metallicity is found to increase
as a function of H band luminosity.  
Spiral galaxies require that both $\tau$ and metallicity 
correlate with H band luminosity: low mass Im+BCD have sub-solar 
$Z$ and $\tau \geq 10$ Gyr, whereas giant spirals have solar 
metallicities and $\tau \sim 3$ Gyr, consistent with elliptical 
galaxies. Moreover we find that the SFH of spiral galaxies in the 
Virgo cluster depends upon 
the presence at their interior of fresh gas capable of sustaining 
the star formation.  In fact the residuals of the $\tau$ vs. $L_H$ 
relation depend significantly on the HI content. HI deficient 
galaxies have shorter (up to a factor of 4) $\tau$ (truncated SFH) 
than spirals with normal HI content.

\end{abstract}




\section{Introduction}

One of the hottest yet unsettled issues of observational cosmology is
the tracing of galaxy evolution over a large interval of
look-back-time, in order to shed light on the processes that led 
to their formation. Currently debated formation models can be broadly
divided into two categories: the hierarchical (Cole et al. 1994) or the
monolithic collapse (Eggen, Lynden-Bell \& Sandage 1962; Sandage
1986) models (see a review by Elmegreen 2001).\\ 
This issue can only now be pursued observationally, with the present 
generation of 10 m telescopes.  Photometric and
spectro-photometric observations of selected samples of galaxies spanning
a large redshift interval can provide us with the time dependence of their
structural parameters, their dynamics, their gaseous content and metal
enrichment (Kauffmann \& Charlot 1998) as well as their star
formation histories.\\ 
An alternative approach to the issue (eg. Bell \& de Jong 2000) is
based on the assumption that galaxies at $z$=0 retain some memory of
their past, which can be unveiled by models (e.g. Bruzual \& Charlot,
1993 population synthesis models).  Observations at $z$=0 would provide
the boundary conditions for the modeling process.\\ 
The fundamental ingredient of both approaches is a robust
observational determination of the "end-point" of galaxy evolution:
i.e. the present stage galaxy properties.  Even this relatively simple 
point is far from settled, since the properties of local galaxies are 
not yet sufficiently known and understood.  For example, it has been 
shown that the present star formation activity (per unit mass)
increases along the Hubble sequence (Roberts\& Haynes 1994; Kennicutt 1998), 
but very little attention has been given to the role of total 
stellar mass in determining the present evolution of
galaxies, an issue that we addressed in Gavazzi et al. (1996a), Gavazzi
\& Scodeggio (1996), Boselli et al. (2001), and we reiterate in the
present work (see also Bell \& de Jong 2000 who stressed the role of 
the central surface brightness in regulating the evolution).\\ 
The properties of the underlying stellar population (and the physical
conditions of the interstellar medium) can be best addressed using
spectro-photometric measurements which are unfortunately only sparsely 
available.
After the pioneering work of Kennicutt (1992), marginal effort was
devoted at deriving the spectral parameters representative of normal
galaxies in the local Universe.  The work of Jansen et al. (2000), who
reports on $\sim$ 200 spectra of local isolated galaxies, represents a
significant exception. In the near future our knowledge of evolved
galaxies will certainly benefit from the SIRTF legacy project SINGS
(http://ircamera.as.arizona.edu/legacy/) by Kennicutt et
al. specifically aimed at obtaining the detailed phenomenology of
local galaxies through all possible observational windows.\\
Another aspect that lately has not been given enough consideration when
studying galaxy formation and evolution is the role of environment in
regulating those processes. 
Clusters of galaxies are "laboratories" where galaxy evolution took
place in significantly different environmental conditions from the
field, the most evident signature of this being the morphology-density
correlation (Dressler 1980), but other more subtle effects are known 
and others wait to be quantified.\\
With the aim of constructing a representative description of galaxies
in a nearby cluster, we undertook a multi-frequency photometric
survey of optically selected galaxies in the Virgo cluster, 
spanning the broadest possible range in morphological type
(Ellipticals, spirals, dE, Im and BCD) and luminosity (-22 $\leq$
$M_B$ $\leq$ -15). Due to its nearness, the Virgo cluster offers
a unique opportunity for carrying out such an analysis, because dwarf
($M_B$ $\leq$ -15) galaxies in this cluster are within reach of 
middle-class telescopes and, owing to the monumental photographic 
work of Sandage, Binggeli \& Tammann which gave origin to the Virgo 
Cluster Catalogue (Binggeli et al. 1985; VCC), for galaxies in this 
cluster we can rely on a particularly accurate morphological 
classification.\\ 
Our own H$\alpha$, optical, near-IR, and millimetric observations,
together with UV (2000 \AA) and centimetric data taken from the
literature, were collected in a multifrequency database.\\
Most recently we undertook a spectro-photometric project aimed at 
obtaining an homogeneous spectroscopic data-set for these galaxies.  
Integrated spectra of 125 Virgo galaxies have been obtained so far, 
covering the spectral range from 3600 to 7000 $\rm \AA$, with a 
resolution of $R$=500-1000. The spectra will be published in a 
forthcoming paper (Paper II of this series).\\
In the present paper we combine the newly obtained spectro-photometric 
data with the photometric observations to map each galaxy spectral energy
distribution (SED) over a very broad wavelength range. These SEDs are
then used with Bruzual \& Charlot (1993, hereafter B\&C) population
synthesis models to constrain the star formation histories of galaxies
in the Virgo cluster.
The sample upon which the present work is based is discussed in
Section 2.  The observations are briefly described in Section 3.  The
construction of the galaxy SEDs, by combining photometry with
spectroscopy, and their corrections are discussed in Section 4. The
method adopted to fit B\&C models to the data is described in Section 5.  
The results of this work are given in Section 6 and discussed in Section 7.

\section {The sample}

\begin{deluxetable}{lc}
\footnotesize
\tablecaption{The completeness in the individual photometric bands
for the 598 Virgo members (or for the 312 spirals) listed in the VCC with $m_p\leq 16$}
\tablewidth{0pt}
\tablehead{
 &  \colhead{N  ~~~~~~~(\%)}}
\startdata
UV         & 134/598  ~(22)   \\ 
U          & 302/598 ~(51)   \\
B          & 393/598 ~(66)   \\
V          & 383/598 ~(64)   \\
J          & 193/598 ~(32)   \\
H          & 355/598 ~(59)   \\
K          & 255/598 ~(43)   \\
(spirals) H$\alpha$  & 228/312 ~(73) \\
(spirals) HI      &  293/312 ~(94)   \\
\enddata
\end{deluxetable}

\begin{deluxetable}{lcccc}
\footnotesize
\tablecaption{The available photometric measurements in the individual bands 
of the spectro-photometric and photometric samples, in three bins of photographic magnitude}
\tablewidth{0pt}
\tablehead{
\colhead{Band} & $m_p$ & \colhead{N Phot~~\%}   &  \colhead{N Phot+Spec~~\%}  & \colhead{N $\Sigma$~~\%}}
\startdata
UV 	 	 & $\leq$16	    &  57/598~(10)   &  63/598~(11) & 120/598 ~(21) \\
"  	 	 & $\leq$15	    &  54/414~(13)   &  62/414~(15) & 116/414 ~(28) \\
"  	 	 & $\leq$14	    &  43/248~(17)   &  50/248~(20) &  93/248 ~(37) \\
		 
U  	 	 & $\leq$16	     &  161/598~(27)   &  105/598~(18) & 266/598 ~(45) \\
"  	 	 & $\leq$15	     &  150/414~(36)   &  94/414~(23) & 244/414 ~(59) \\
"  	 	 & $\leq$14	     &  125/248~(50)   &  77/248~(31) & 202/248 ~(81) \\
		 
B~or~V	 	 & $\leq$16	&  206/598~(34)   &  124/598~(21) & 330/598 ~(55) \\
"	         & $\leq$15     &  178/414~(43)   &  109/414~(26) & 287/414 ~(69) \\
"      	         & $\leq$14     &  136/248~(55)   &  86/248~(35) & 222/248 ~(90) \\
		 
J	 	 & $\leq$16	     &  118/598~(20)   &  64/598~(11) & 182/598 ~(31) \\
"	 	 & $\leq$15	     &  105/414~(25)   &  57/414~(14) & 162/414 ~(39) \\
"	 	 & $\leq$14	     &   89/248~(36)   &  51/248~(21) & 140/248 ~(57) \\
		 
H	 	 & $\leq$16	     &  191/598~(32)   &  100/598~(17) & 291/598 ~(49) \\
"	 	 & $\leq$15	     &  171/414~(41)   &  91/414~(22) & 262/414 ~(63) \\
'	 	 & $\leq$14	     &  134/248~(54)   &  77/248~(31) & 211/248 ~(85) \\
		 
K	 	 & $\leq$16	     &  136/598~(23)   &  101/598~(17) & 237/598 ~(40) \\
"	 	 & $\leq$15	     &  117/414~(28)   &  87/414~(21) & 204/414 ~(49) \\
"	 	 & $\leq$14	     &   96/248~(39)   &  72/248~(29) & 168/248 ~(68) \\
		 
spir.~HI 	 & $\leq$16   &  114/312~(37)	&  86/312~(28) & 200/312 ~(65) \\
"	 	 & $\leq$15   &   98/234~(42)	&  77/234~(33) & 175/234 ~(75) \\
"	 	 & $\leq$14   &  78/150~(52)   &  60/150~(40) & 138/150 ~(92) \\
		 
spir.~H$\alpha$ & $\leq$16  &  92/312~(29)   &  87/312~(28) & 179/312 ~(57) \\
"		 & $\leq$15  &  78/234~(33)   &  78/234~(33) & 156/234 ~(66) \\
"		 & $\leq$14  &  62/150~(41)   &  60/150~(40) & 122/150 ~(81) \\
\enddata
\end{deluxetable}

\begin{deluxetable}{llccccccc}
\footnotesize
\tablecaption{The completeness of the spectro-photometric and photometric samples
in three bins of photographic magnitude.}
\tablewidth{0pt}
\tablehead{
\colhead{mag. limit} & \colhead{N VCC} & \colhead{with z}   & \colhead{N Phot}   & 
\colhead{(\%)} & \colhead{N Phot+Spec}  & \colhead{(\%)} & \colhead{N $\Sigma$}  & \colhead{(\%)}}
\startdata
(incomplete)     &        &        &  220  &   -   &  125 &   -   &  345 & -  \\
$\leq$16         &  598   & 552    &  206  &  (34) &  124 & (21)  &  330 &  (55) \\
$\leq$15         &  414   & 411    &  178  &  (43) &  109 & (26)  &  287 &  (69) \\
$\leq$14         &  248   & 248    &  136  &  (55) &   86 & (35)  &  222 &  (90) \\
$\theta < 2 ~or~ 4 < \theta < 6 ~from M87$ &       & & & &        &	 &   \\
(incomplete)     &        &        &   85  &   -   &  108 &   -   &  193 &  - \\
$\leq$16         &  270   & 244    &   73  &  (27) &  108 & (40)  &  181 &  (67) \\
$\leq$15         &  185   & 184    &   53  &  (29) &   95 & (51)  &  148 &  (80) \\
$\leq$14         &  107   & 107    &   28  &  (26) &   75 & (70)  &  103 &  (96) \\
\enddata
\end{deluxetable}

\begin{deluxetable}{cccccc}
\footnotesize
\tablecaption{The new spectra in the Virgo Sample.}
\tablewidth{0pt}
\tablehead{
\colhead{Obs. run} & \colhead{Tel.} & \colhead{Istr.} & \colhead{Resolution} & \colhead{Slit Width}& \colhead{N. Obs.}\\
\colhead{} &\colhead{} & \colhead{} & \colhead{} & \colhead{arcsec}& \colhead{}}
\startdata
OHP 1998  &1.93m & Carelec	   &  500 & 2.5 &   10\\
OHP 1999  &1.93m & Carelec	   &  1000 & 2.5 &  17\\
OHP 2000  &1.93m & Carelec	   &  1000 & 2.5 &  26\\
OHP 2001  &1.93m & Carelec	   &  1000 & 2.5 &  24\\
ESO 2001  & 3.6m & EFOSC2	   &  500 & 1.5 &   39\\
WHT 2001  & 4.2m & ISIS 	   &  2000 & 2.5 &  7 \\
\enddata
\end{deluxetable}

\begin{figure}[!t]
\epsscale{0.6} 
\centerline{\plotone{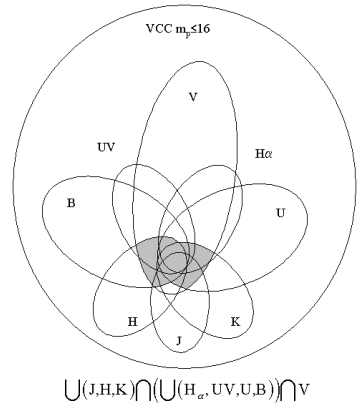}}
\small{\caption{330 objects meet the condition for inclusion in the 
"photometric sample" (with $m_p\leq 16$). Of these 124, observed spectroscopically, constitute 
the "spectro-photometric sample".}\label{insiemi}}
\end{figure}

\begin{deluxetable}{llrr}
\footnotesize
\tablecaption{The linear regression analysis}
\tablewidth{0pt}
\tablehead{
\colhead{sample}    & \colhead{regression}   & \colhead{R} & 
\colhead{see Fig.}}
\startdata
Ellipticals         &  $\rm Log \tau= -0.086 \times LogL_H  + 1.351 $     &  -0.537 &  \ref{Taulum_ell}\\ 
Ellipticals         &  $\rm B-H= 0.302 \times LogL_H  + 0.599 $     &  0.711 &  \ref{colmag}\\
Spirals             &  $\rm Log \tau= -0.149 \times LogL_H  + 2.221$      &  -0.604 &  \ref{Taulum_spir}\\ 
Spirals             &  $\rm \Delta Log \tau = -0.132 \times Def_{HI}  + 0.071$      &  -0.529 &  \ref{resTaulum_spir}\\ 
\enddata
\end{deluxetable}
The Virgo Cluster Catalogue (VCC, Binggeli et al. 1985) lists 598 
bona-fide Virgo cluster members brighter than $m_p$ = 16.0. 
Of these, 552 are members spectroscopically confirmed by Binggeli et 
al. (1993) or by Gavazzi et al. (2000c), while the remaining 46 are
classified as "possible cluster members". Among the 598
galaxies, 312 are of late-type (from Sa to Im-BCD).\\
The availability of photometric data
for these VCC galaxies is detailed in Table 1. \\
For the purpose of this paper we selected, among Virgo members with 
available photometry, a subsample of 330 objects (only 4 are not 
spectroscopically confirmed members) according to the following criteria:\\
1) They have the V magnitude plus at least one
NIR magnitude (J or H or K; H in most cases) and at least one blue/UV
magnitude (B or U or UV or H$\alpha$) measurement available.
Fig.\ref{insiemi} provides a graphical representation of this
selection criterion.\\
2) To avoid large errors on total magnitudes that would derive from
the extrapolation of poorly defined growth curves, we require B, V and 
NIR magnitudes to be obtained from at least 3 aperture measurements (only 5
galaxies were rejected because of this requirement. This same
requirement does not apply to UV magnitudes because these are already 
total magnitudes).\\
Among these 330 objects, 124 galaxies with available spectra form the
"spectro-photometric" sample. Galaxies meeting criteria 1) and 2), but
without spectroscopic observations, constitute our "photometric 
sample".\\ 
Details of the sample completeness in three bins of photographic magnitude
\begin{figure}[!t]
\epsscale{1.0} 
\centerline{\plotone{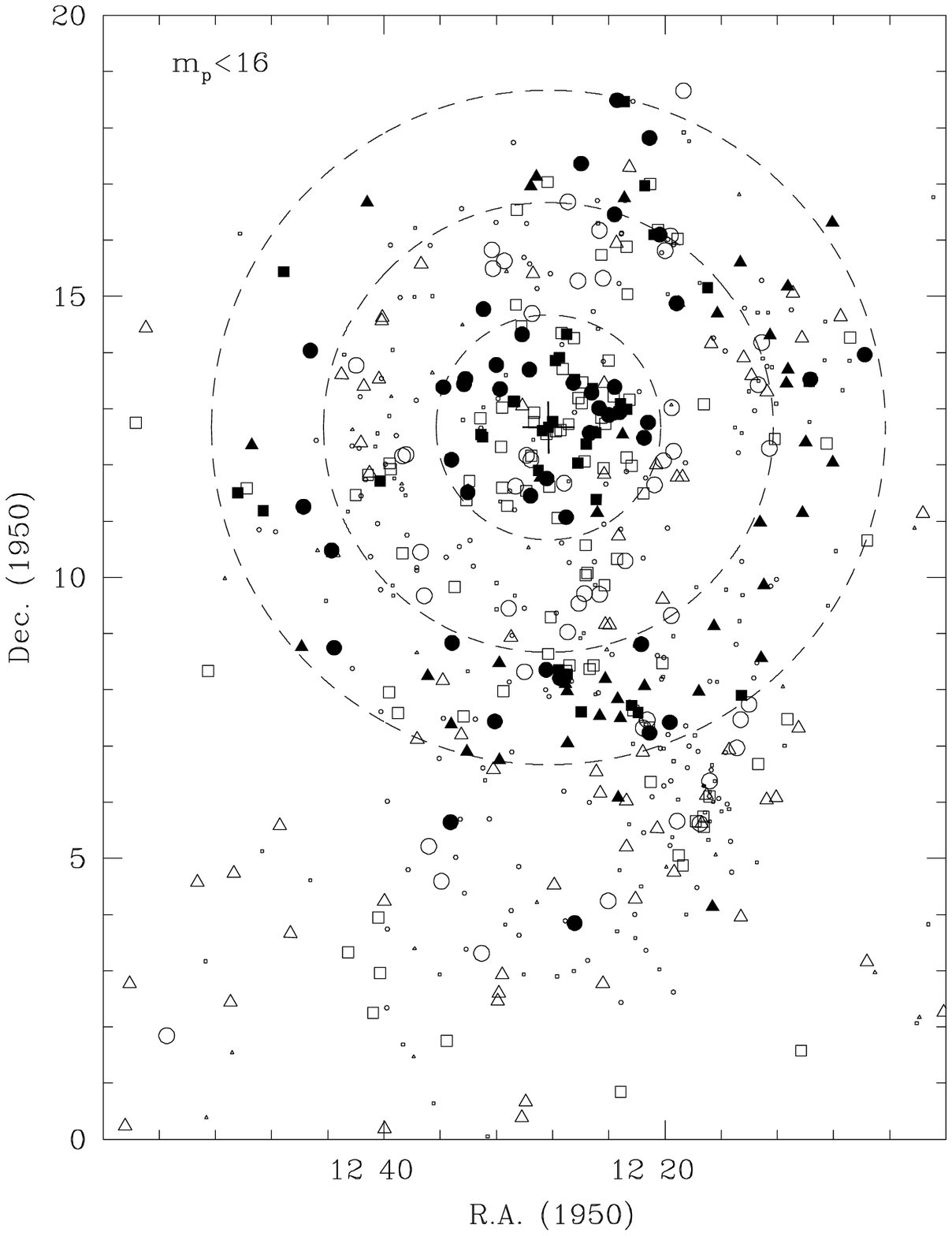}}
\small{\caption{Sky distribution of the 598 galaxies brighter than
$m_p \leq 16.0$ in the VCC. The filled symbols represent galaxies in
the spectro-photometric sample, the empty ones galaxies in the
photometric sample, while the tiny symbols refer to unobserved
galaxies.  Elliptical galaxies are given with squares, spiral galaxies
with $\rm Def_{HI}<0.5$ are triangles, spiral galaxies with $\rm Def_{HI}>0.5$
are circles. Circles are drawn at 2, 4, 6 deg. projected radial
distance from M87 (cross).}\label{celestial}}
\end{figure} 
and for the individual photometric bands are given separately for the
"photometric" and "spectro-photometric" samples, and for their sum 
in Table 2, while the final samples and their completeness are 
illustrated in Table 3.
Their sky distribution is shown in Fig. \ref{celestial}.
While the global completeness of the "spectro-photometric sample" is
still rather poor at $m_p$ $\leq$ 16.0 (20\%), it improves
significantly (to 40\%) over the restricted area composed by the
annuli with angular distance $\theta<2$~deg and $4\leq \theta \leq 6$
from M87 where we concentrated most of our spectroscopic effort
(108 spectra).\\
In addition to the 330 galaxies selected with the above criteria we
include in our analysis 15 objects with $m_p$ $\geq$ 16.0 (mostly
dwarf ellipticals and BCDs), out of which 14 have photometry and one 
(VCC636) has a spectrum available.  These are: VCC 328, 636, 793, 810, 
872, 882, 916, 1065, 1173, 1313, 1352, 1353, 1377, 1420, 1577.
This addition brings to 345 and 125 the number of galaxies in the 
"photometric" and "spectro-photometric" samples, respectively 
(as listed in the first line of Table 3).\\ 
Following Gavazzi et al. (1999b) we assume a distance of 17 Mpc for the
members (and possible members) of Virgo cluster A, 22 Mpc for Virgo
cluster B, 32 Mpc for objects in the M and W clouds, adopting $H_o$ =
75 $\rm km~s^{-1} Mpc^{-1}$.\\

\section {The Data}

\subsection{Spectro-photometry}
Spectro-photometric measurements of galaxies in the Virgo cluster were
taken during several runs from 1998 to 2001, using various
telescopes. The detailed presentation of the spectra, including the
line analysis, will be addressed in a forthcoming paper (Gavazzi et
al. Paper II, in preparation).  Here we give some concise information
focused on the continuum spectral properties.  Out of the 125 spectra
used in this work, 2 were taken from the catalogue of Kennicutt
(1992). Spectra of 7 BCD galaxies, obtained with the William Herschel
Telescope with a resolution of 2000, were kindly provided to us by 
J.M. Vilchez. All the remaining spectra
were obtained by us: 77 using the OHP 1.93m and 39 using the ESO 3.6m
telescope (see Table 4).\\ 
The observations were taken in "drift mode": {\it i.e.} with the slit,
generally parallel to the galaxy major axis, drifting over most of the
optical surface of the galaxy (as in Kennicutt 1992).  
Spectra taken in this way are representative of the entire galaxy, 
and not only of its central regions.  All spectra cover the wavelength 
range 3600-7000 \AA~ with a resolution of 500 (ESO) and 1000 (OHP). 
Spectra were flux-calibrated using several spectrophotometric 
standards.  However, because of the use of drift-scan mode, their 
absolute flux calibration is meaningless, and therefore all spectra 
were normalized to their value at $\lambda=5500~$\AA.\\

\subsection {Photometry}

Total UV, optical (U,B,V) and NIR (J,H,K) magnitudes are used to
complement the spectroscopic observations in the construction of the
galaxies SEDs on which this work is based.\\
U, B and V  photometry is generally derived from our own CCD 
measurements. When these are not available it is derived
from aperture photometry taken from the Longo et
al. (1983) catalogue and the LEDA database (Prugnel \& Heauderau 
1998). We used only galaxies with at least three independent aperture 
measurements, after checking for foreground star contamination.
We compute total magnitudes $\rm (U,B,V)_{25}$, i.e. magnitudes 
computed at the optical radius (as in Gavazzi \& Boselli 1996), 
which are, on average, 0.1 mag fainter than the total asymptotic 
magnitudes.\\
NIR data from Nicmos3 observations, obtained by us, have been
presented in Gavazzi et al. (1996b,c), Gavazzi et al. (2000a), Gavazzi
et al. (2001), Boselli et al. (1997), Boselli et al. (2000).  In most
cases we measured the H band magnitude, in many cases the K band one, 
in a few cases the J band one. From these data we derive total 
magnitudes $\rm H_{25}$, determined as described in Gavazzi \& 
Boselli (1996). When $\rm H_{25}$ is not available we derive
it from $\rm K_{25}$ using $\rm H-K$=0.25 mag. $\rm H_{25}$ values 
are converted into total luminosities using: 
$\rm log L_H = 11.36 - 0.4 H_{25} +2LogD$ (in solar
units), where D is the distance to the source (in Mpc).
The assumed photometrical uncertainties are 15 \% for B, V and H and 
20 \% for U, J and K (see also appendix C).\\
For 120 of the 345 galaxies in the photometric sample 
(63 in the spectro-photometric sample) UV data are available 
from three balloon experiments:
the FOCA (Milliard et al. 1991),
FAUST (Lampton et al. 1993) and SCAP (Donas et al. 1987). SCAP and 
FOCA use 150 \AA~ wide filters centered at 2000 \AA, while FAUST is 
centered at 1650 \AA ~($\Delta \lambda = 250 \AA$).
The published FOCA magnitudes (Donas et al. 1991, 1995), were 
recently corrected (Donas, private communication) to account for 
better cross-calibration between the various experiments. 
These are total (not aperture) UV magnitudes extracted from the 
photographic plates. The FAUST data were converted by us to 2000 \AA~ 
(using 0.2 mag correction on average, as concluded by Deharveng et al. 1994).
The quoted error on the UV magnitude is 0.3 mag in general, but it 
ranges from 0.2 mag for bright galaxies to 0.5 mag for weak sources 
observed in frames with larger than average calibration 
uncertainties. We assume a conservative 0.4 mag uncertainty for 
all galaxies to account for all sources of systematic errors.

\subsection{Other data}

H$\alpha$+[NII] fluxes enter indirectly in the SEDs determination,
as they provide an estimate of the ionizing flux below 912 \AA~ 
(see Section 4.1 and Appendix A). They were measured in our spectra, when
available, or derived from imaging or aperture photometry by Kennicutt
\& Kent (1983), Almoznino \& Brosch (1998),
Heller et al. (1999), Koopmann et al. (2001) and references therein.
Additional imaging observations of 150 galaxies have been 
recently obtained by us during several runs at the Observatoire 
d'Haute Provence (France), at Calar Alto (Spain) (Boselli \& Gavazzi 
2002), at San Pedro Martir (Mexico) (Gavazzi et al. 2002) and at 
the INT (Boselli et al. 2002).  The estimated error
on the H$\alpha$+[NII] flux is $\sim$ 20\%.\\ 
Another important ingredient in this work, although it does not enter
into the SEDs determination, is the estimate of a galaxy current neutral 
hydrogen content. HI data are taken from Hoffman et al. (1996, and 
references therein). HI fluxes are transformed into neutral hydrogen 
masses with an uncertainty of $\sim$ 10\%. 
From the hydrogen mass the HI deficiency parameter 
($\rm Def_{HI}$) is computed according to Giovanelli \& Haynes (1985): 
$\rm Def_{HI}=<log M_{HI}(T^{obs}, D^{obs}_{opt})> - log M^{obs}_{HI}$
where the observed HI mass is compared with the value expected from
an isolated (i.e. free from external influences) galaxy having the same
morphological type $\rm T^{obs}$ and optical linear diameter 
$\rm D^{obs}_{opt}$ (for details, see Haynes \& Giovanelli 1984).

\section {The galaxy SEDs}

Out of the 125 obtained spectra, 119 (2 Seyfert or Liners and 4 
unclassified galaxies are not included) are given in
Fig. \ref{template_spectra}, grouped in 10 bins of Hubble type.  These
template spectra of normal cluster galaxies allow us to trace the
dependence of the mean spectral properties along the Hubble sequence.
Red continua characterize early type galaxies up to Sab-Sb (included),
while the continua become progressively bluer for later types.  Except
for dEs, which have weaker absorption lines, also the line properties 
of galaxies up to Sab-Sb appear indistinguishable, including the 4000
\AA~ break.  Emission lines are absent among galaxies earlier than
Sa. Sa and Sab-Sb show no lines other than H$\alpha$ and [NII].  For
later types the emission lines become progressively stronger,
including [OIII], [OII] and the other Balmer lines. The accurate analysis of
the line properties is postponed to Paper II of this series.\\
The method we use to constrain a galaxy star formation history
consists of running B\&C models with a broad grid of model parameters,
and fitting the synthesized spectra to the observed ones to derive a
set of best-fit parameters. However, given the limited wavelength
coverage of the optical spectroscopy, a similar approach would not
sufficiently narrow down the parameter space.  Much stronger
constraints can be derived if the spectroscopic data are combined with
photometry taken over the broadest possible wavelength baseline.  We
thus combine our 125 spectra (``spectro-photometric sample'') with UV,
optical (UBV) and infrared (JHK) photometric data.  For this purpose
magnitudes are converted into fluxes, then normalized at 5500 \AA,
as for the spectra. Before the combination, both spectroscopic and
photometric data are corrected for reddening according to the
prescriptions of Appendix B.  In this way we obtain SEDs that cover
the domain 2000--22000 \AA. For  220 objects in the ``photometric
sample'' that are not part of the ``spectro-photometric'' one the B\&C
models are fitted to the photometric data alone, as in Bell \& de Jong
(2000).\\
In order to extend our SEDs below 2000 \AA~ we include the estimate of
the ionizing flux below
\begin{onecolumn}
\begin{figure}
\epsscale{1.0} \plotone{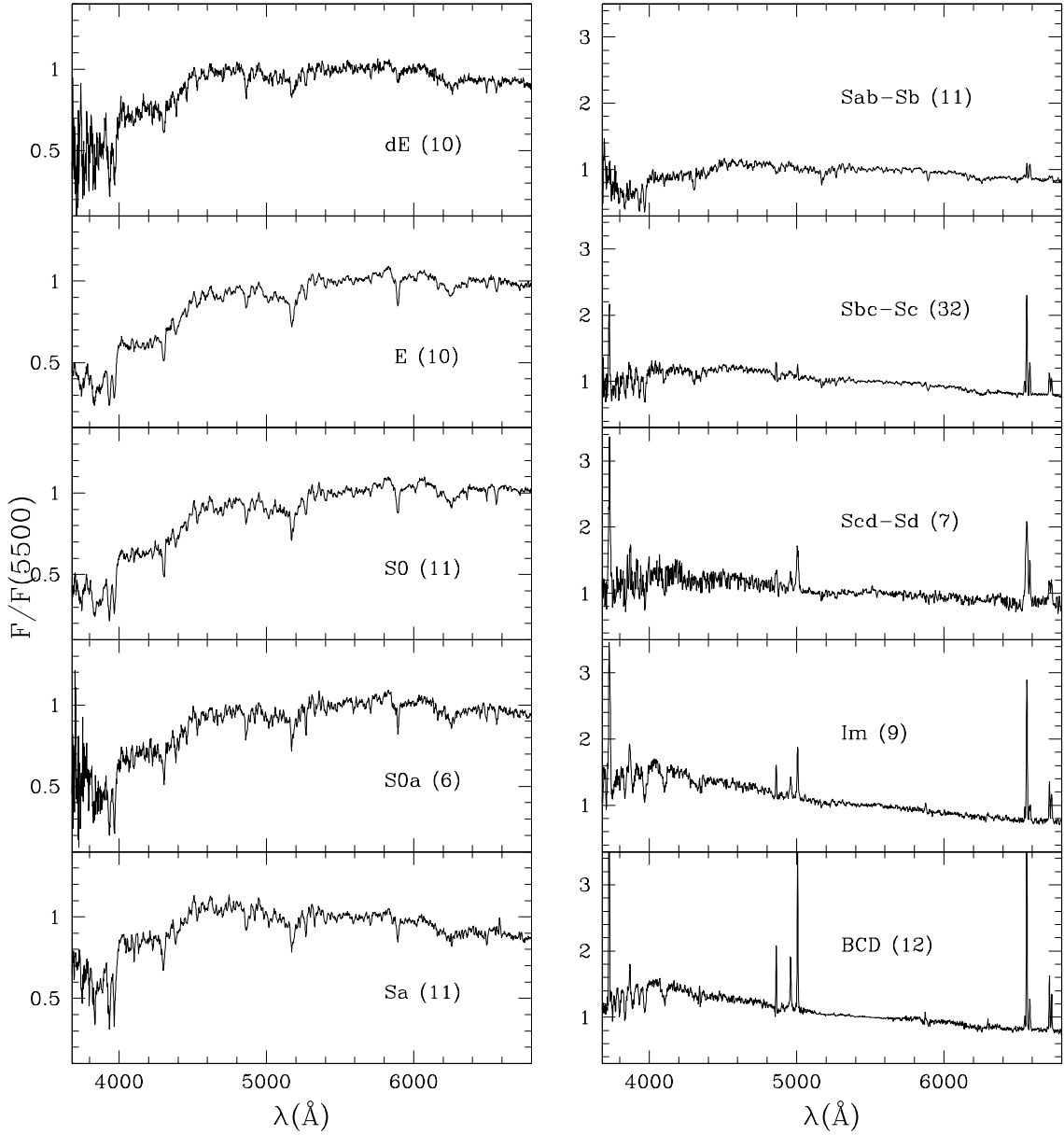} 
\small{\caption{Template optical spectra in 10 bins of Hubble type.
The number of averaged spectra in each bin is given in parenthesis.
Note that the flux scale differs between the left and the right
panels.}\label{template_spectra}}
\end{figure}
\end{onecolumn}
\begin{twocolumn}
912 \AA, as obtained from the H$\alpha$ flux.
This is based on the assumption that the flux in the Balmer line
H$\alpha$ is due to recombination of hydrogen atoms within the HII
regions excited by $<912$ \AA~ photons from the stellar radiation
field.  The details of this calculation are given in Appendix A.\\
An example of such an extended composite SED (for the galaxy VCC 1205)
is shown in Fig.\ref{Ttdegeneracy}. The dots with error bars represent
the available broad-band photometry with its uncertainty.  The optical
spectrum is represented by the thick continuous line. The dashed
horizontal line in the far UV represents the $<912$ \AA~ flux estimate
derived from H$\alpha$. The thin continuous lines represent the
Bruzual \& Charlot models fitted to the data (see the next section).

\section{The Bruzual \& Charlot models}

We use the 2001 version of Bruzual \& Charlot (1993) population
synthesis models. These models provide the time ($T$) evolution of the
synthesized spectra of galaxies characterized by an initial
metallicity ($Z$), a Star Formation History (SFH) and an Initial Mass
Function (IMF):\\
\begin{equation} 
F_\lambda(T)[Z,SFH,IMF]
\end{equation}
Two star formation histories are considered:\\ 
a) the exponential SFH:\\
\begin{equation} 
SFR(T_{exp},\tau_{exp})= 1/\tau~exp(-T/\tau) 
\end{equation} 
which describes the exponential time evolution of a burst at $T$=0.\\
b) the "delayed-exponential" SFH or "a la Sandage":
\begin{equation} 
SFR(T_{San},\tau_{San})= T/\tau^2~exp(-T^2/2\tau^2)
\end{equation}
which mimics the SFH first proposed by Sandage (1986) (see
Fig. \ref{sandage}).  The temporal evolution of this family of
functions is a delayed rise of the SFR up to a maximum, followed by an
exponential decrease. Both the delay time and the steepness of the
decay are regulated by a single parameter $\tau$.\\ 
We always assume a Salpeter IMF ($\alpha$ = 2.35 from 0.1 to 100
M$\odot$; Salpeter 1955), and explore a parameter grid in $Z$ and
$\tau$. $Z$ is let free to vary from 1/50 to 2.5 Z$\odot$ in 5 steps:
0.0004, 0.004, 0.008, 0.02, 0.05. 
These values correspond to the initial metallicity grid computed 
by B\&C and we have not
tried to interpolate between them.
\begin{figure}[!t]
\epsscale{1.}
\plotone{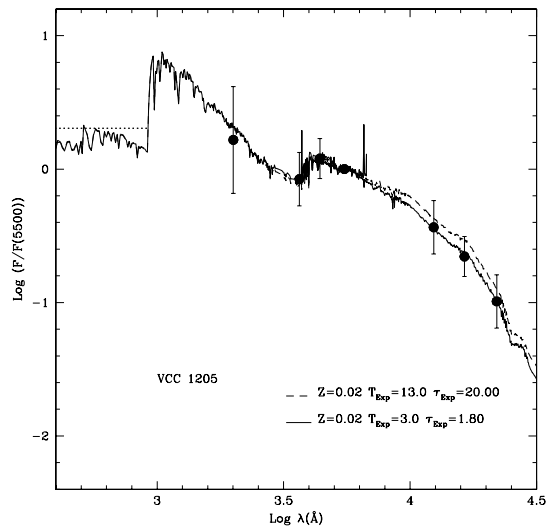}
\small{\caption{The extended SED of the Sc galaxy VCC 1205: the dots
with error bars represent the available broad-band photometry with its
uncertainty, while the observed optical spectrum is represented by the
thick continuous line. The dashed horizontal line in the far UV
represents the $<912$ \AA~ flux estimate derived from H$\alpha$. The
SED is fitted with B\&C models with exponential SFH and solar
metallicity (thin continuous and dashed lines). In spite of having
chosen two very different combinations of ages and $\tau$ ($T$=3 with
$\tau$=1.8 and $T$=13 with $\tau$=20 Gyr) the two solutions have similar
probabilities.  This example illustrates the degeneracy between $T$ and
$\tau$.}\label{Ttdegeneracy}}
\end{figure}
$\tau$ varies from 0.1 to 25 Gyr in 45 approximately logarithmic steps. 
All models are computed at two epochs: $T$=5, 13 Gyr.\\
Fig. \ref{Models} shows a sample of the synthetic galaxy spectra
obtained assuming $T$=13 Gyr and  $T_{San},\tau_{San}$, our 
preferred choice for these two options (see next Section). 
In the top panel we explore the effects of
changing the time-scale of the star formation process $\tau$ while
keeping the metallicity fixed at the solar value. A broad range of
spectral properties is obtained, most noticeably when $\tau<5$ Gyr. 
In the bottom panel we keep $\tau$ fixed at 3.5 Gyr, and let $Z$ vary 
over its full parameter range. In this case a much narrower
spectral variety is obtained.\\ 
The fitting of these synthetic spectra to the
\begin{figure}[!t]
\epsscale{1.}
\plotone{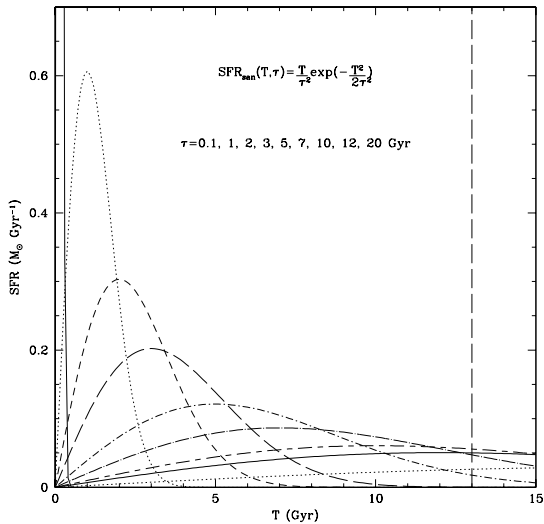} 
\small{\caption{The time evolution of a star formation event "a la Sandage" 
is plotted for 9 values of $\tau$ from its origin ($T$=0) to
15 Gyr.  This SFH is characterized by a steep rise followed by an
exponential decay. It becomes progressively shallower and the position
of the peak is delayed in time with increasing $\tau$. At the assumed
age of 13 Gyr (vertical line) the SFH is rising for $\tau>13$
Gyr.}\label{sandage}}
\end{figure}
observed galaxy SED is
carried out with a chi-squared minimization procedure that identifies
the $Z$ and $\tau$ parameters for the most likely matching synthetic
SED. Sometimes there is no clear minimum in the distribution of 
chi-squared values, and the values flatten out until one reaches 
the edges of the parameter grid. In this case we assign lower or 
upper limits (as appropriate in each single case) to the values 
of the corresponding parameters. Details of the fitting procedures 
are presented in Appendix C.

\section{Results}

\subsection{$T$ vs. $\tau$ degeneracy}

Fig. \ref{Ttdegeneracy} shows the main limitation of our method as an
age estimator: the SED of VCC 1205 is fitted with two B\&C models with
$T_{exp},\tau_{exp}$, solar metallicity and with very different 
ages ($T$=3, 13 Gyr) and decay times ($\tau$=1.8, 20 Gyr), both models 
giving results consistent with the observations.
\begin{figure}[!t]
\epsscale{1.}
\plotone{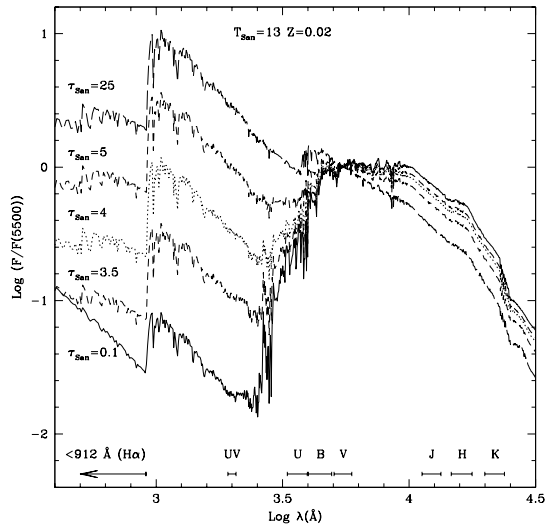}
\plotone{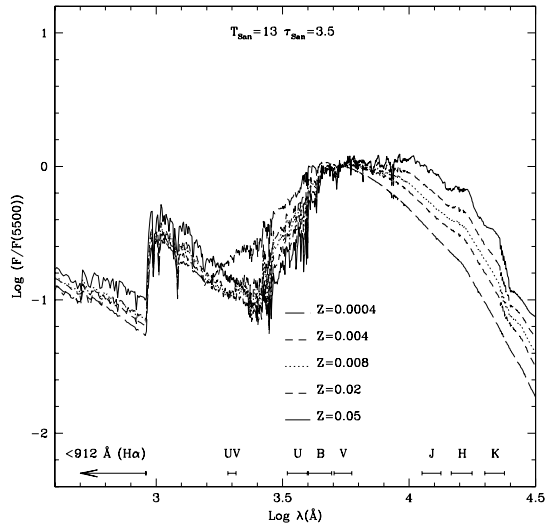}
\small{\caption{Synthetic B\&C galaxy spectra for a galaxy age $T$=13
Gyr and a star formation history  ``a la Sandage''. In the top panel
the models have fixed initial metallicity ($Z$=0.02) and cover a 
range of $\tau$ values, while in the bottom panel the models have 
fixed $\tau=3.5$ Gyr and cover a range of $Z$ values.}\label{Models}}
\end{figure}
Therefore, beside the
well known age-metallicity degeneracy for a single stellar population 
(Worthey 1994), there exists another type of degeneracy that can 
significantly affect our attempts at deriving an accurate 
reconstruction of a galaxy star formation history. In fact, because 
of the existing strong correlation between T and $\tau$, in spite of 
the broad wavelength coverage of our data, our method cannot 
disentangle the system age from $\tau$, but it provides only $\tau$/T.\\
This is further illustrated in Fig.\ref{tsutau} where the ratios
$\tau_{San}/T_{San}$ computed for two different ages ($T$=5 and 13
Gyr) are plotted one against the other, showing that $\tau/T$ is 
approximately constant while $T$ varies by more than a factor of two.
Due to this intrinsic weakness of the method we are forced to run B\&C
models with T fixed arbitrarily. We have chosen to use $T$=5 and 13 Gyr,
but hereafter we show results only for $T$=13 Gyr, because the
assumption of a common age of 5 Gyr for all galaxies appears to be at
odds with the very existence of high redshift galaxies.


\subsection{The choice of SFH}

Having set $T$=13 Gyr we show in Fig.\ref{color_color} a comparison
between the observed color-color ($U-V$ vs. $B-H$) relation for the
galaxies in our sample and the grid of model colors obtained with the
two adopted SFH (Exp, San). While both SFH reproduce without problems
the reddest observed colors, the exponential SFH fails to reproduce
the colors of the bluest galaxies, reaching at best a color $U-V$ $\sim$
0.5 mag redder than observed, even
assuming $\tau$=20 Gyr (quasi continuum SFH).  The Sandage SFH does a
better job at reproducing the full range of observed colors within the
explored range of $\tau$ and metallicity, since it includes rising SFR
at $T$=13 Gyr, thus allowing bluer colors. For this reason, and because
it offers a more realistic time evolution of the galactic star
formation history, we adopt hereafter the Sandage SFH as the best
single parameterization for the star formation history of galaxies of
all types.
By allowing $\tau$ to assume negative values the exponential 
law can also have a rising SFR, but this solution is less elegant 
than the Sandage law. Another way to create these blue colors is 
through starbursts and episodic star formation, which can have 
considerable effect especially on small galaxies.

\subsection{Fit to the template SEDs}

As discussed in the previous two sections, we select the Sandage SFH
as the simplest and perhaps most realistic representation of the 
SFH of
\begin{figure}[!t]
\epsscale{1.}
\plotone{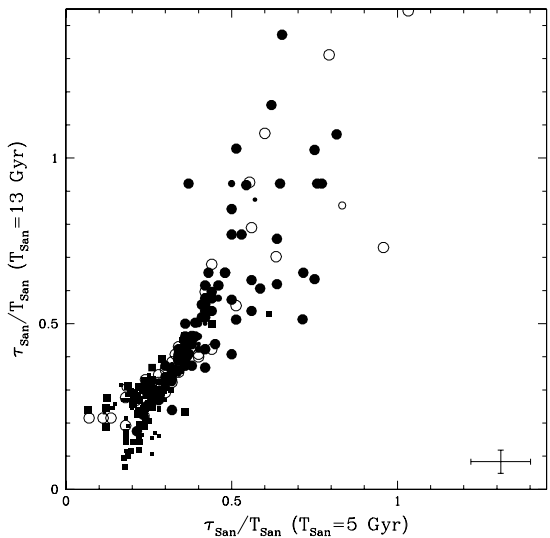} 
\small{\caption{The value of the ratio $\tau_{San}$/$T_{San}$ for the 
best fitting model with $T$=5 Gyr against that for the best fitting 
model with $T$=13 Gyr, further illustrating the degeneracy between 
$T$ and $\tau$.
Symbol shape represents Hubble type: elliptical galaxies are
represented with squares, spirals with circles. Open and filled
symbols differentiate galaxies in terms of HI deficiency (only for
spirals): open symbols for galaxies with $\rm Def_{HI}>0.5$ while filled
symbols for galaxies with $\rm Def_{HI}<0.5$. Symbol size provides an
indication of the best fitting model metallicity (size is decreasing
with increasing metallicity). The typical uncertainty is given in the 
corner of the figure.}\label{tsutau}}
\end{figure}
galaxies consistent with
a fixed age of $T$=13 Gyr, and we let $Z$ and $\tau$ vary as free
parameters in our fitting procedure.\\
First we apply the fitting procedure to our templates SEDs. These were
obtained extending with photometric data the spectral
templates shown in Fig. \ref{template_spectra}, and are therefore very
robust determinations of the typical SED for each Hubble type. The
results of the fit are shown in Fig. \ref{template_fit}.
Good quality fits are obtained
with $\tau$ increasing approximately 
monotonically along the Hubble
sequence. It appears that the whole Hubble sequence can be modeled
assuming increasingly delayed star formation histories of longer
duration. Early type galaxies (dE, E, S0) have $\tau \leq 3$ Gyr,
Sa--Sc have intermediate $4<\tau<6$ Gyr, with uncertainties $\sim 1$ Gyr.
Galaxies in the 3 latest bins of Hubble
\begin{figure}[!t]
\epsscale{1.}
\plotone{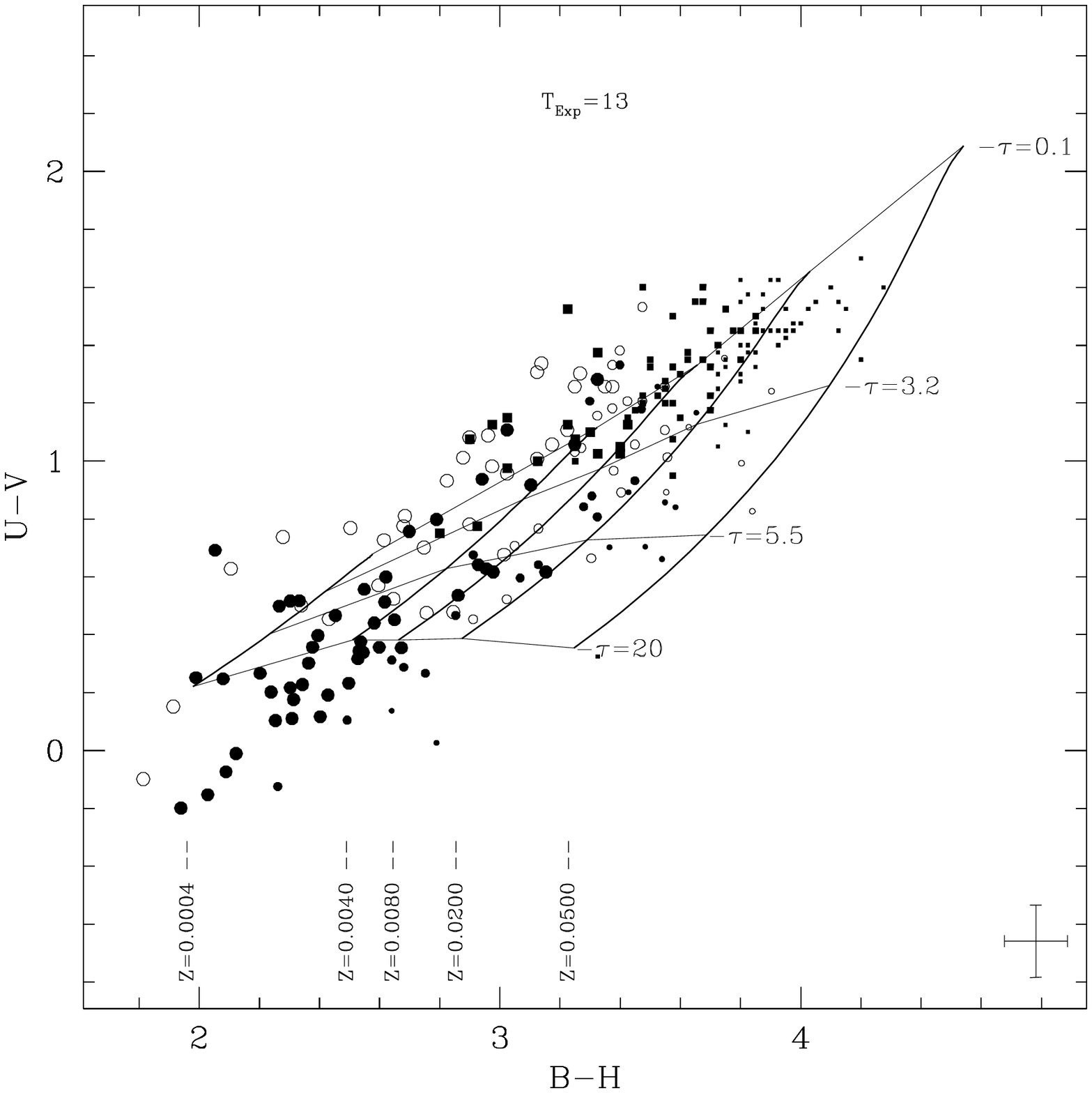}
\plotone{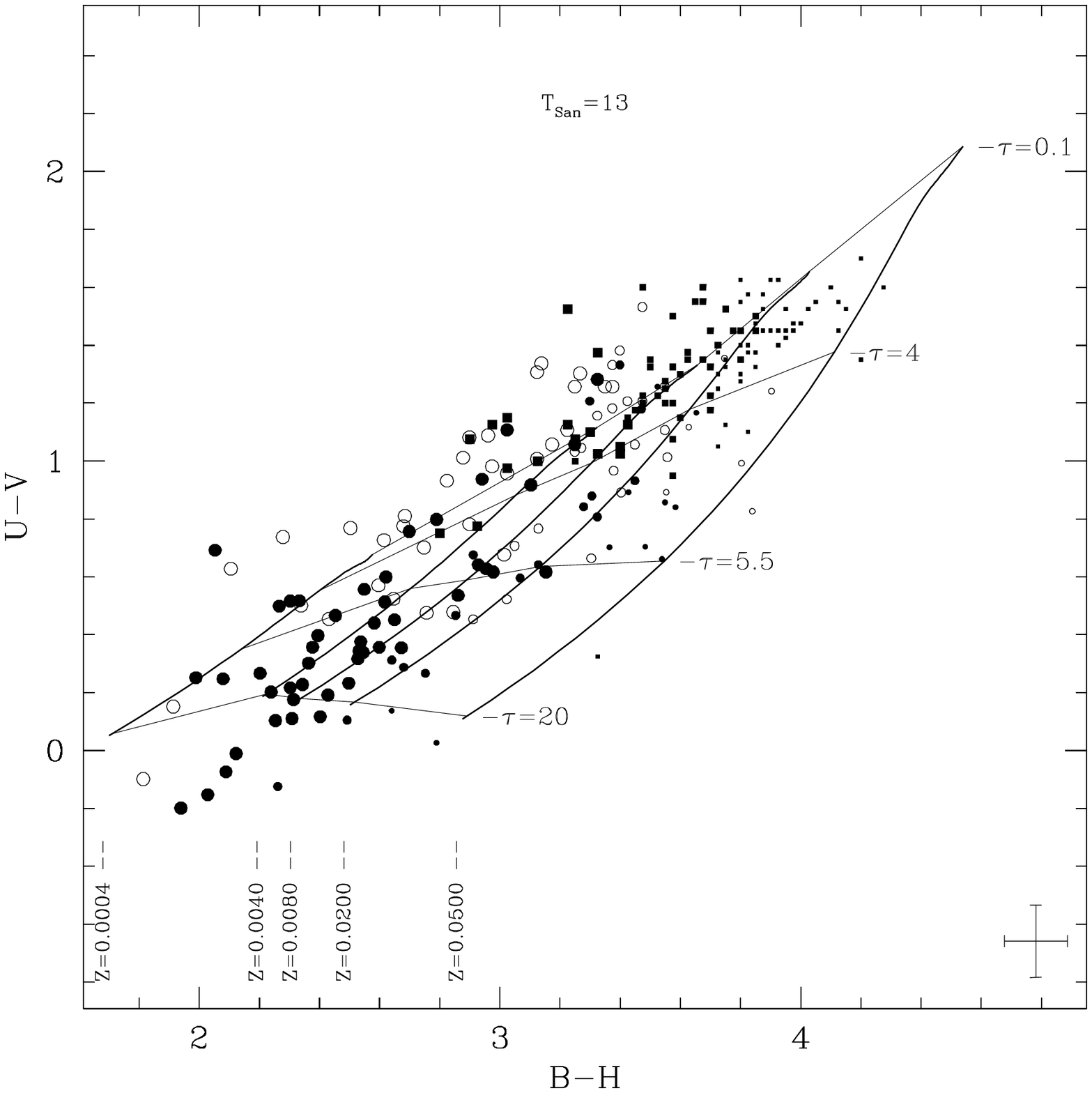}
\small{\caption{The observed (corrected for internal extinction) 
$U-V$ vs. $B-H$ color-color distribution is
compared with the corresponding color-color relation obtained from the
B\&C models (lines) for two star formation histories: the exponential
one (top) and the one  ``a la Sandage'' (bottom) (same symbols as in
Fig.\ref{tsutau}).} \label{color_color}}
\end{figure}
type are consistent with 
constant or even rising SFHs. Im and BCD have lower limits 
$\tau \geq 7$ Gyr. The behavior of the metallicity along the Hubble
sequence is more erratic. Approximately solar metallicities are found 
mostly in early-type objects, except for dEs which show a sub-solar 
metallicity.  Slightly sub-solar metallicities are found across
the entire range of Hubble types. Beside dEs, also Im+BCD galaxies 
show sub-solar metallicities.  
Fig. \ref{template_contour} shows the iso-confidence contours from the
fitting procedure as a function of $\tau$ and $Z$. Contours are
plotted at 0.68, 0.85, 0.99 probability. The figure shows how $\tau$ is 
better constrained than $Z$. Typical errors on $\tau$ (see Sa, Sb, Sc) 
are 0.9 Gyr. The uncertainty on the metallicity is typically 
$\sim 0.012$, i.e. approximately one step of the metallicity grid.
 
\subsection{The SFH of elliptical galaxies as a function of  luminosity}

The fit to template SEDs discussed in the previous section has
shown a significant variation of the fit parameters along the Hubble
sequence.  In this and in the following section we apply the fitting
procedure to the individual SEDs, separately for Early and Late type
galaxies. The fits were inspected one by one, but due to their large
number, they are not shown individually. \\
We explore the dependence of the fit parameters on the NIR luminosity
$L_H$, which traces the bulk of the stellar mass. This tight
correlation between stellar mass and NIR luminosity has been discussed
by Gavazzi et al. (1996a) and Pierini et al. (2002) for luminous
spiral galaxies, where HI mass does not dominate over the stellar
mass, and by Zibetti et al. (2002) for elliptical galaxies. 
Thus the analysis presented in this and in the following 
section should shed some light on how the SFH of galaxies varies with 
their systemic mass.\\ 
We reiterate that individual elliptical galaxies in the "photometric"
and "spectro-photometric" samples are fitted with B\&C models with
fixed SFH (Sandage) and $T$=13 Gyr and the dependence of the derived
parameters $\tau$ and $Z$ are analyzed as a function of the system
luminosity $L_H$.  Fig.\ref{Taulum_ell} shows the shallow dependence
of $\tau$ on the luminosity $L_H$ for elliptical galaxies (the linear
regression of this and of the following relations are given in Table
5).  For a range of $L_H$ spanning 4 decades, $\tau$ varies between 2
and 4 Gyr: i.e. these galaxies have experienced a "short" burst of
star formation in their early history (see Fig.\ref{sandage}), with
giant ellipticals being only twice "earlier" than dEs.
\end{twocolumn}
\begin{onecolumn}
\begin{figure}
\epsscale{1.}
\plotone{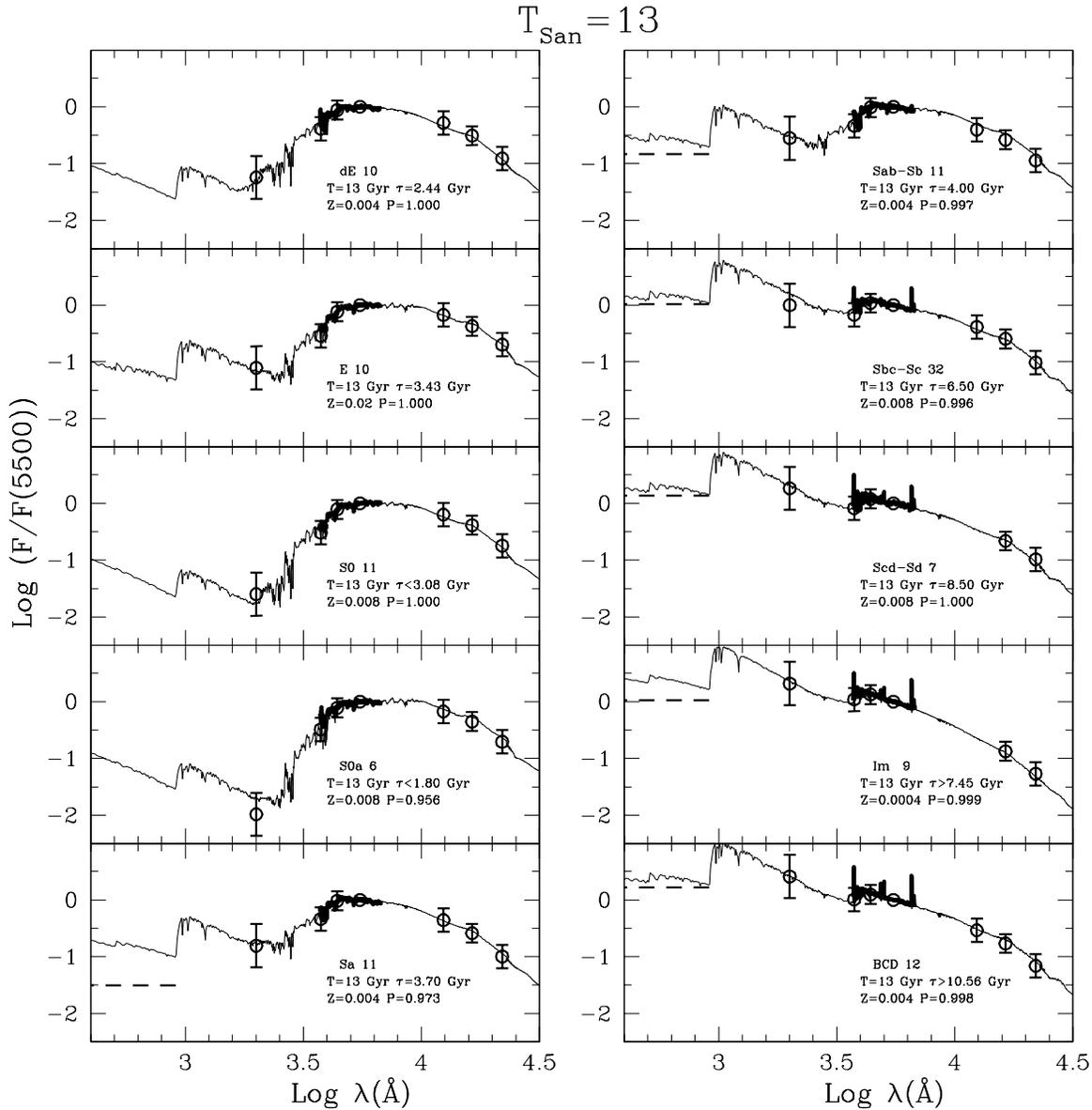}
\small{\caption{The (corrected for internal extinction) template SEDs , obtained averaging the photometry
and the individual spectra grouped in 10 bins of Hubble type (same as
Fig. \ref{template_spectra}) are fitted with B\&C models assuming the
Sandage SFH at $T$=13 Gyr, letting $Z$ and $\tau$ as free parameters
(given in the labels together with the fit probability).  
}\label{template_fit}}
\end{figure}
\end{onecolumn}
\begin{twocolumn}
\begin{figure}[!t]
\epsscale{1.1}
\plotone{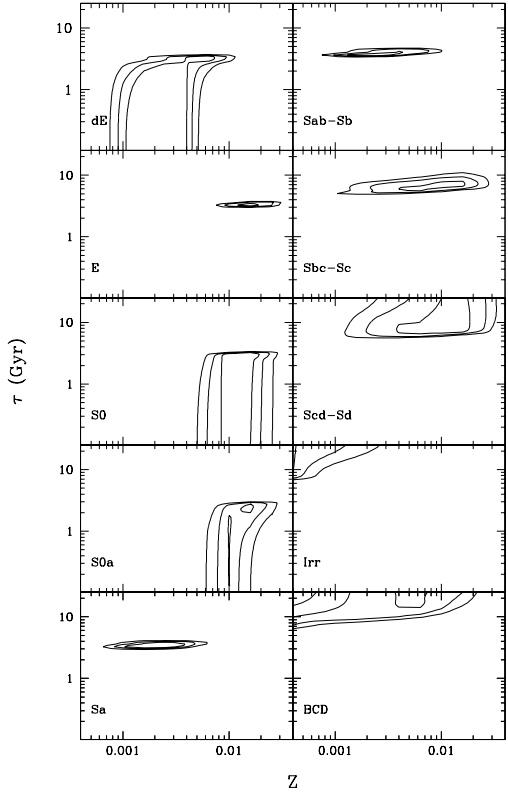}
\small{\caption{The confidence intervals as function of $\tau$ and $Z$ 
for the 10 Hubble type classes (same as
Fig. \ref{template_fit}) are given with contours drawn at 
0.68, 0.85, 0.99 probability.}\label{template_contour}} 
\end{figure}
The symbol size in Fig.\ref{Taulum_ell} increases with decreasing
metallicity of the best fitting model. This helps showing that, beside
the shallow $\tau$ vs. $L_H$ relation, there is a more pronounced
dependence of the model initial metallicity on the system luminosity, 
as shown in Fig.\ref{zlum_ell}.  
At any given $L_H$, lower metallicity systems seem to have a 
lower $\tau$, due to the age-metallicity degeneracy.
Summarizing, elliptical galaxies have
experienced consistently "short" bursts of star formation "early" in
their history. Their metallicity increases from sub-solar
(1/5 Z$\odot$) for the dwarfs to solar for the giants.
\begin{figure}[!t]
\epsscale{1.}
\plotone{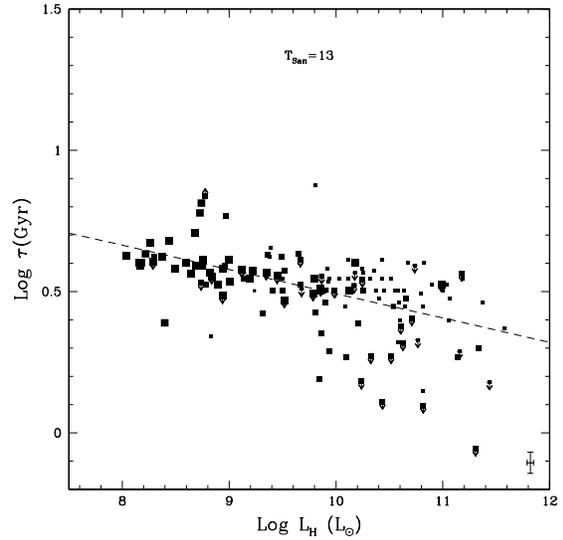}
\small{\caption{The dependence of the fitted $\tau$ on H band
luminosity for elliptical galaxies (same symbols as in
Fig.\ref{tsutau}). Arrows mark upper and lower
limits (see Appendix C).}\label{Taulum_ell}}
\end{figure}
\begin{figure}[!t]
\epsscale{1.}
\plotone{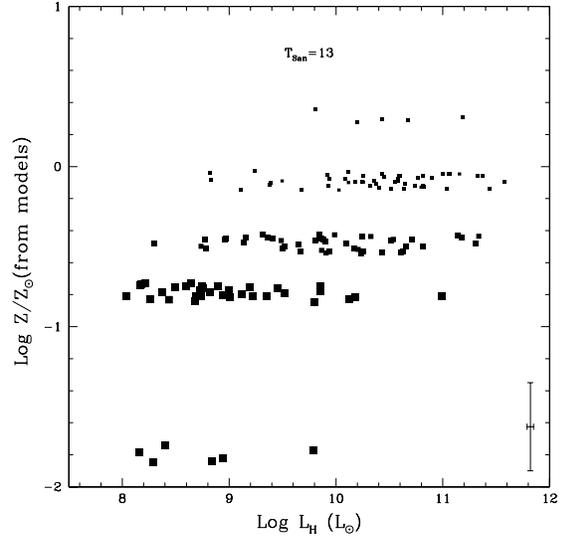}
\small{\caption{The dependence of the fitted initial metallicity on H band
luminosity for elliptical galaxies (same symbols as in
Fig.\ref{tsutau}). A small random number has been added
to the metallicity in order to avoid superposition of
points.}\label{zlum_ell}} 
\end{figure}

\subsection{The SFH of spiral galaxies as a function of luminosity}

Figure \ref{Taulum_spir} shows that the dependence of $\tau$ on the
system luminosity $L_H$ is significantly steeper for spiral than for
elliptical galaxies.  Dwarf Irrs have $\tau$ of 10 or more Gyr
(increasing SFR), while the most massive spirals have $\tau$ of
approximately 3 Gyr, consistent with elliptical galaxies.
\begin{figure}[!t]
\epsscale{1.} 
\plotone{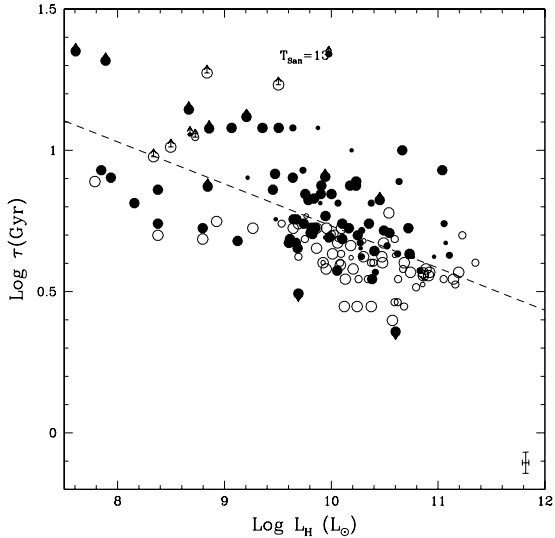} 
\small{\caption{The dependence of the fitted $\tau$ on H band
luminosity for spiral galaxies (same symbols as in Fig.\ref{tsutau}). 
Arrows mark upper and lower limits (see Appendix C).}\label{Taulum_spir}}
\end{figure}
\begin{figure}[!t]
\epsscale{1.} 
\plotone{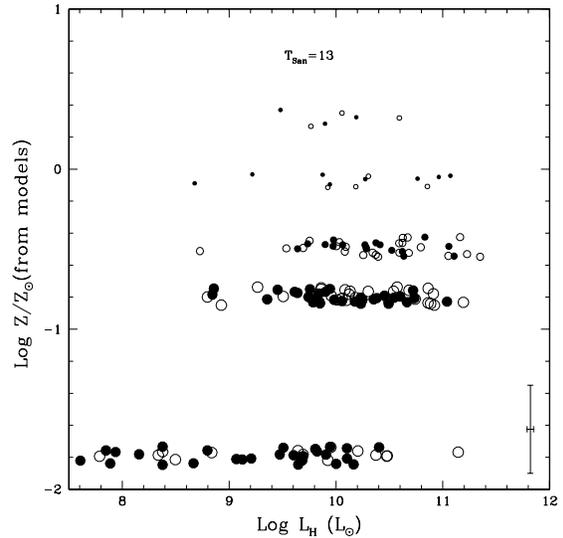}
\small{\caption{The dependence of the fitted metallicity on H band
luminosity for spiral galaxies (same symbols as in Fig.\ref{tsutau}). 
A small random number has been added to the
metallicity in order to avoid superposition of
points.}\label{zlum_spir}}
\end{figure}
Once again in this figure the symbol size is inversely proportional to 
the metallicity of the best fitting model.  While we observe a 
lower mean metallicity for any range of $L_H$ than in the case of 
elliptical galaxies, spirals have a global dependence of the model 
metallicity on $L_H$ not too different from the one discussed above for 
the ellipticals, as shown in Fig.\ref{zlum_spir}.
This dependence of metallicity on $L_H$ confirms earlier findings by 
Bell \& de Jong (2000). 
However, as far as $\tau$ is concerned, these authors
claim that the principal correlation is the one between $\tau$ and 
the central K band surface brightness, followed by a less significant 
$\tau$ vs. $L_K$ relation. Our data show a marginally better 
correlation of $\tau$ vs. $L_H$ than that of $\tau$ vs. $\mu_e$. 
Notice however that we use the effective H band surface brightness $\mu_e$ 
instead of the central one used by Bell \& de Jong (2000), and this 
might contribute to the discrepancy. 
We repeated the residual test of the Bell \& de Jong and
we find that the residuals of the $\tau$ vs. $\mu_e$ relation
correlate with $L_H$  slightly better than the residuals of the 
$\tau$ vs. $L_H$ relation correlate with $\mu_e$. Based on our 
analysis alone it is therefore impossible to discriminate whether 
$\mu_e$ or $L_H$ is the principal parameter in regulating galaxy 
star formation history.

\subsection{The color-magnitude relation}

The color ($B-H$) vs. magnitude ($L_H$) relation for all galaxies,
irrespective of their morphological type, is given in
Fig.\ref{colmag}.  
Elliptical galaxies obey to a shallow color-magnitude relation
which can be understood almost exclusively in terms of increasing
metallicity with mass (the metallicity of dEs is significantly lower
than that of giant ellipticals), with only a marginal spread in $\tau$
in the explored luminosity range.  
The steeper (and non-linear) color-magnitude sequence of 
spirals/Im+BCD, instead, derives from the combination of increasing
$Z$ and decreasing $\tau$ from dwarfs to giants.  Notice that the colors
of ellipticals are as red as those of spirals at the highest
luminosities, suggesting a smooth transition from elliptical to spiral 
galaxies, with S0 galaxies in between.
\begin{figure}[!t]
\epsscale{1.}
\plotone{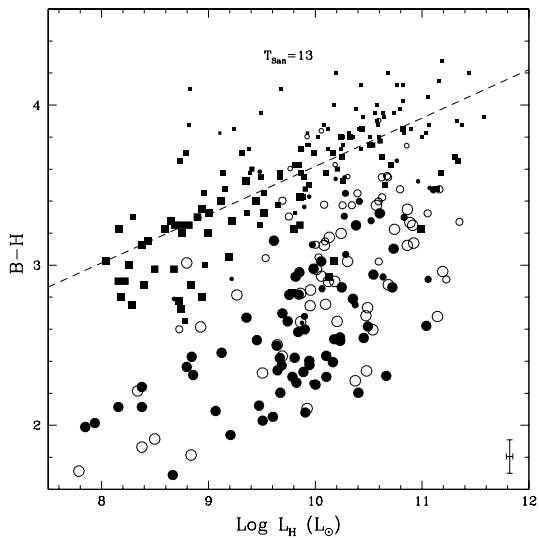}
\small{\caption{The (corrected) color--magnitude ($B-H$ vs. $L_H$) 
relation for elliptical and spiral galaxies (same symbols as in 
Fig.\ref{tsutau}). The dashed line is the fit for the elliptical 
galaxies. }\label{colmag}}
\end{figure}

\subsection{The environmental dependence of the SFH of spiral galaxies}

The results of the previous sections are probably not unique to the
Virgo cluster.  They are consistent with Bell \& de Jong (2000)
analysis of non-cluster galaxies and it would not be surprising to see
similar luminosity dependencies among the isolated galaxies of Jansen
et al. (2000).  Are there features depending on the specific cluster
environment?  
Ideally one should investigate whether the $\tau$ - luminosity relation
varies as a function of some environmental parameter, like the local galaxy
density or the projected angular distance from the cluster center.
However Virgo has a complex structure, reflecting its non-virialized nature,
which makes this approach unfeasible.
Beside the main cluster A containing M87, a number of clouds are known 
to exist in this cluster: cluster B dominated by M49, clouds W, M, E 
and S, each with different velocity or distance, or both, as analyzed 
by Gavazzi et al. (1999). As a result of the superposition of 
clouds on any line of sight, both the local galaxy
density or the projected angular distance from M87 provide
little indication about the real environment surrounding any given
galaxy in the Virgo cluster.\\
We are therefore using a slightly indirect approach to the analysis of
environmental effects, based on the 
$\rm Def_{HI}$ parameter, as defined by Giovanelli \& Haynes (1985)
(see also Sect. 3.3). Analyzing HI data of spiral galaxies in 9 rich clusters
of galaxies, these authors concluded that HI deficiency is most likely
due to ram-pressure stripping, a
dynamical phenomenon occurring to galaxies in their fast motion
through the cluster intergalactic medium (Gunn \& Gott 1972).
The extension of the sample, from the original 9 to 18 clusters,
allowed Solanes at al. (2001) to conclude that the most HI deficient objects 
are those in radial orbits. The fact that 
asymmetries in the HI distribution (Dickey \& Gavazzi 1991,
Bravo-Alfaro et al. 2000) are often associated with radio continuum
trails (Gavazzi \& Jaffe 1987, Gavazzi et al. 1995) and in one case 
with an H$\alpha$ trail (Gavazzi et al. 2001b) 
pointing in the same direction as the HI asymmetry, strongly suggests 
that ram-pressure stripping is the most plausible explanation for 
the pattern of HI deficiency in clusters. Thus the $\rm Def_{HI}$ 
parameter can be considered an indirect indicator of the environmental
conditions experimented by a galaxy over the last few Gyr.
A careful inspection of Fig.\ref{Taulum_spir} shows that
galaxies suffering from significant HI deficiency ($\rm Def_{HI}>$0.5)
(empty circles) are found systematically below the regression line of the
$\tau$ - luminosity relation.  The residual of the $\tau$ -
luminosity relation versus $\rm Def_{HI}$ is plotted in
Fig.\ref{resTaulum_spir} showing a significant trend.
The deficient galaxies have SFH $\sim$4 times shorter than the unperturbed
galaxies.
Summarizing: spiral galaxies have luminosity dependent SFHs, 
and the residual of the $\tau$ - luminosity relation is
found to depend on the HI content, with the gas-poor galaxies showing
truncated SFH compared with cluster galaxies with normal HI content.

\section{Discussion and conclusion}

We have combined spectro-photometry for 125 galaxies in the Virgo
cluster with broad-band photometry from UV to NIR (including an
estimate of the $<912$ \AA~ flux from H$\alpha$). The resulting SEDs
(corrected for internal extinction) were fitted with Bruzual \&
Charlot population synthesis models with a variety of model parameters
($SFH$, $T$, $\tau$, $Z$).  The main results of the present work can be
summarized as follows:\\ 
\begin{figure}[!t]
\epsscale{1.}
\plotone{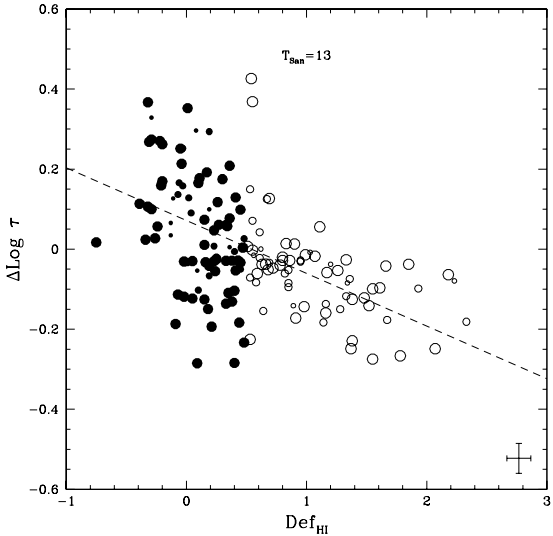}
\small{\caption{The residual of the relation between $\tau$ and H band
luminosity for spiral galaxies (Fig.\ref{Taulum_spir}) is given as a
function of the HI deficiency parameter (same symbols as in
Fig.\ref{tsutau}).}\label{resTaulum_spir}}
\end{figure}
1) the observations are consistent with the simplest (minimum number
of assumptions) hypothesis that the all galaxies began their formation
13 Gyr ago.  This assumption implies that the star formation history
of galaxies is allowed to increase with time, i.e adopting a SFH 
"a la Sandage" (see Fig.\ref{sandage}). According to this family of SFHs
the SFR increases with time, reaches a maximum and decreases
exponentially.  A single parameter $\tau$ determines the delay of the
SFR peak, thus the mean "age" of the stellar population, as well as
the shape of its time dependence: the shorter $\tau$ the more peaked
the SFHs.\\ 
Having made these assumptions the Hubble sequence is obtained with
increasing $\tau$ with lateness, as proposed by Sandage (1986).\\ 
2) $\tau$ is a decreasing function of the  H-band luminosity.  This
dependence is shallow for elliptical galaxies: i.e. the value of 
$\tau$ for the dwarfs ($\tau \sim 4$ Gyr) and that for the giants 
($\tau \sim 2$ Gyr) do not differ by more than a factor of 2, 
both being short compared with the assumed galaxy age 
(see Fig.\ref{Taulum_ell}).  Late-type galaxies display a broader 
range of $\tau$ values, from 25 Gyr (increasing SFR) for the low mass 
dwarf Im+BCD galaxies, down to 3 Gyr for the massive, early type spirals
(see Fig.\ref{Taulum_spir}).\\ 
3) the stellar metallicity increases as a function of the  H-band 
luminosity across the entire Hubble sequence (see Figs.\ref{zlum_ell} and
\ref{zlum_spir}).  $Z$ is found to span approximately a decade from low
mass to massive systems. It should however be stressed that 
metallicities are poorly constrained by the model.\\ 
In short: the galaxy mass, as estimated from the H-band luminosity (or
perhaps the surface brightness), seems to play a crucial role in determining
most galaxy structural parameters (similar dependences of several
other photometric and structural parameters on H-band luminosity
are discussed in Scodeggio et al. 2002).\\ 
Trying to reconcile the above results with models of galaxy
formation-evolution is beyond the scope of the present investigation.
We refer the reader to Elmegreen (2001) for a comparison of the
predictions of the hierarchical and of the monolithic models with the
observations, unfortunately not explicitly focused on the dependence
of the model predictions on the galaxy mass, an issue that we would
like to address in the remainder of the present discussion, at least
qualitatively.\\ 
The hierarchical scenario contains an implicit reference to the
mass. The mass in this model is a growing function of time (i.e. of
the number of mergings). It is natural that the most massive objects
(either giant ellipticals or bulge-dominated spirals), being the
product of a small number of major merging events which were likely to
take place at $z>2$ (Lacey \& Cole 1993), can be modeled in our
formalism with $\tau \sim 3$ Gyr.  Low mass Im+BCD with significant
current SFR, can be reconciled with the hierarchical scenario because
dwarf galaxies either presently in formation or currently undergoing minor
mergings are allowed by the model. Their SFH, which we model for
simplicity with shallow, continuous functions, may in fact consist of
a series of minor star formation bursts, associated with minor merging
events (see Kauffmann, Charlot \& Balogh 2001).\\ 
Our findings can in principle be reconciled also with the monolithic
scenario if the mechanism of proto-galactic collapse is assumed to be
not scale-free: i.e. that it depends on the total mass of the
proto-galactic cloud, beside the dependence on its angular momentum.  
Angular momentum
would discriminate between the formation of ellipticals and spirals
with identical mass (Sandage 1986), whereas the system mass would
further regulate the efficiency of collapse: massive proto-galaxies
would undergo an early efficient collapse, while small primordial
fluctuations would take a longer time to form galaxies. As discussed
by Boselli et al. (2001), not only the star formation history, but
also the time dependence of the gas consumption can be modeled within
the monolithic scenario with the above assumption.  \\
The additional finding of the present work is that the star
formation time-scale $\tau$ of spiral galaxies in the Virgo cluster
depends on their present hydrogen content.  
Highly HI deficient spirals in Virgo 
have typical time-scales ($\tau$) up to 4 times smaller
than their HI healthy counterparts.
If, as proposed by Giovanelli \& Haynes (1985) and confirmed by 
Solanes et al. (2001), HI deficiency occurs to cluster spirals due to 
ram-pressure stripping (Gunn \& Gott 1972), the exhaustion of the HI reservoir
might have produced an earlier truncation of the
star formation activity in these galaxies.
This mechanism is invoked by Gavazzi et al. (2002) to explain the 
significantly quenched current star formation rate found in HI 
deficient Virgo spirals included in their extensive H$\alpha$ imaging 
survey. One might argue that the
argument could be reversed: these galaxies appear HI deficient because
a higher gas consumption occurred during an intense earlier star
formation burst.  However this reversed argument is in contradiction
with the observations because it would imply that HI deficient
galaxies would exist also outside clusters, something that is not
observed.\\
Among the highly HI deficient objects with truncated SFH there are few
bright Sa in our sample (VCC958 = NGC4419, VCC984 = NGC4425, VCC1047 =
NGC4440, VCC1158 = NGC4461, VCC1412 = NGC4503) with $\tau$ as short as
3 Gyr.  This look-back time corresponds to $z \sim 3$, suggesting that
at this early cosmological epoch Virgo was a fully developed cluster
with a dense IGM.  This is uncomfortably close to the maximum time of
cluster formation allowed by the hierarchical theory ($z \sim 2$).

\begin{acknowledgements}

We wish to thank J. Donas for providing us with unpublished UV magnitudes;
James Lequeux, Jean Michel Deharveng, Veronique Buat
for interesting discussions, Stephan Charlot for his comments and for 
providing us with the
latest version of the stellar population synthesis models. An unknown 
referee is acknowledged for his constructive criticism.\\
\end{acknowledgements}



\end{twocolumn}

\appendix

\section{Estimate of the $<912$ \AA~ flux from $H\alpha+[NII]$.}

The stellar radiation field with $\lambda<912$ \AA~ ionizes the gas 
which re-emits, via recombination lines.
If the gas is optically thick in the Lyman continuum, the number
of photons in a specific recombination line is directly proportional
to the number of star photons in the Lyman continuum.
In the case of H$\beta$ this number is given by
equation (5.23) in Osterbrock (1989).
For H$\alpha$ we have:
\begin{equation}
\int\limits_{\nu_{0}}^{\infty }\frac{L_\nu}{h\nu}d\nu=L_{H\alpha}\cdot C
\end{equation}
where:
\begin{equation}\label{uvha}
1/C=h\nu_{H{\beta}}\frac{\alpha^{eff}_{H\beta}(H^o,T)}{\alpha_B(H^o,T)} \frac{F_{H\alpha}}{F_{H\beta}}
\end{equation}
Assuming $T$=10000K and the Osterbrock case B:\\
$\alpha^{eff}_{H\beta}(H^o,T)=3\times 10^{-14} ({\rm cm^3~sec^{-1}})$\\
$\alpha_B(H^o,T)=2.59\times 10^{-13}({\rm cm^3~sec^{-1}})$\\
and
$\frac{F_{H\alpha}}{F_{H\beta}}=2.87$\\
From the H$_\alpha$ luminosity it is thus possible to recover the number of ionizing photons, 
which can be compared with the similar quantity derived from the integral on the model spectrum.\\
For galaxies in the "spectro-photometric" sample $L_{H_\alpha}$ is measured. 
For galaxies in the "photometric" sample $L_{H_\alpha}$ can be in most cases derived from
imaging observations. These however must first be corrected for the contamination of [NII].
From our spectra we calibrate an empirical relation between ${H\alpha}/[NII]$ and $L_H$
and apply it to estimate [NII] for the remaining galaxies of known $L_H$.\\
The derived $L_{H\alpha}$ must be further corrected for extinction.
We assume 0.8 and 0.6 mag extinction at H$\alpha$ for galaxies earlier than Scd
and for Sd-Im-BDC respectively (consistently with Boselli et al. 2001).
Moreover we assume that 100\% of the UV photons contribute to excite the recombination lines 
(Mezger 1978), with en escaping fraction of 0.
A conservative estimate of the uncertainty on the derived $<912$ \AA~ flux is 1 mag. 
 
 \section{The extinction correction}

\begin{figure}[!t]
\epsscale{0.85}
\plotone{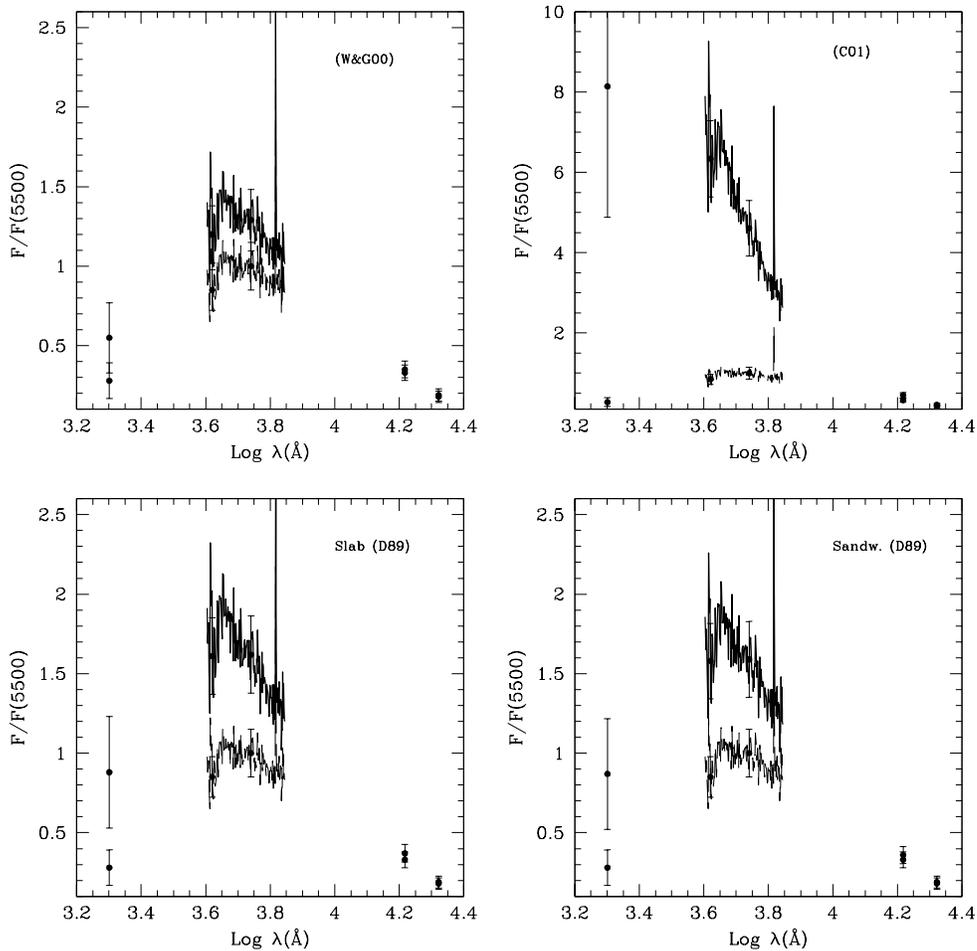}
\small{\caption{The SED of VCC2058 is shown uncorrected (light curve) and corrected according to
the dusty-spherical slab model of Witt \& Gordon (2000) (top-left panel); Calzetti (2001) (top-right panel);
the slab model of Disney et al. (1989)  (bottom-left panel) and the sandwich model of Disney et al. (1989)
(bottom-right panel).}\label{VCC2058}}
\end{figure}

The issue of determining the internal extinction in galaxies is a very 
controversial (from the extreme optically transparent assumption of Holmberg, 1958
to the optically thick case of Valentijn 1990) and difficult one,
due to the complexity of the mutual geometry of obscuring dust and stars (Charlot \& Fall 2000). 
Moreover stars of different ages have different scale heights, making the determination  
of the wavelength dependence of extinction a very uncertain task.
The scale height ratio dust/stars was found to vary with $\lambda$ 
from 0.2 in H to 0.9 in B by Boselli \& Gavazzi (1994), whereas 
Xilouris et al. 1999 found it approximately constant.\\
Solving this issue is beyond the scope of the present investigation.
However we cannot avoid correcting our extended SEDs for internal 
extinction using some, even simplistic recipe.\\
The extinction corrected intensity at any $\lambda$ is related to 
the observed intensity by:
\begin{equation}
LogI_{\lambda C}= LogI_{\lambda o} + 0.4 \cdot A_\lambda
\end{equation}
where 
\begin{equation}
A_\lambda = f(\tau_\lambda)
\end{equation}
$\tau(\lambda)$ is the optical thickness at $\lambda$ and $f$ depends on the relative geometry
of dust and stars.\\
We tested several approaches, whose results are shown in Fig. \ref{VCC2058}, carrying 
the SED of the moderately inclined Sc galaxy VCC 2058, 
before and after the correction for internal extinction.\\
1) We tried the Calzetti (2001, C01 hereafter) empirical absorption law derived for
star-burst galaxies from the Balmer decrement (top-right panel of Fig. \ref{VCC2058}). 
As pointed out by Buat et al. (2002) the Calzetti
law can hardly apply to normal galaxies, as it overestimates the absorption in the UV, 
due to the particular geometry of star-burst galaxies.\\
2) We derived $A_{UV}= g(FIR/UV)$ from the FIR/UV ratio, 
as proposed by Buat et al. (1999, 2002)
and Witt \& Gordon (2000; W\&G00 hereafter). This choice is supported by the fact that FIR and UV fluxes are
in first approximation independent from the SFH (the two are produced by similar stellar populations)
and from the assumed dust-star geometry. 
From $A_{UV}$ we calculate $\tau_{UV}$ for 3 geometries, than we derive
\begin{equation}
\tau_\lambda= f'(A_\lambda) 
\end{equation} 
\begin{equation}
\tau_\lambda = \tau_{UV} \cdot k_\lambda / k_{UV}
\end{equation}
using the galactic extinction law $k(\lambda)$ (Savage \& Mathis 1979),
and we compute the complete set of $A_\lambda$ using (B2).\\
The assumed geometries are:\\
2a) the sandwich model of Disney et al. (1989, D89 hereafter) 
(bottom-right panel of Fig. \ref{VCC2058}). We used a $\lambda$ dependent dust to stars scale ratio: 
0.86(B), 0.78(V), 0.18(H) (intermediate between the optically thick and thin case of Boselli \& Gavazzi 1994).\\
2b) the slab model of D89 (bottom-left panel of Fig. \ref{VCC2058}).\\
2c) the dusty-spherical slab model (with clumpy dust distribution) of W\&G00, 
which includes the contribution from the scattered light 
(top-left panel of Fig. \ref{VCC2058}).\\
As concluded earlier, we rule out the Calzetti law because it overestimates 
$A_{UV}$ in normal galaxies.
Similarly we excluded the W\&G00 model because the spherical approximation is hardly
a realistic one.  Since the sandwich and slab models of D89 produce very consistent results
we finally adopted for simplicity the slab model.\\
In this model the explicit functional dependence of (B2) gives the face-on:
\begin{equation}
A_\lambda = 2.5 \cdot Log \frac{1-exp(-\tau_\lambda \cdot sec(i))}{\tau_\lambda\cdot sec(i)}
\end{equation}
where $sec(i)=a/b$.  
According to Buat et al. (1999) $A_{UV}$ can be quantified as:
\begin{equation}
A_{UV} (mag) = 0.466 + Log(FIR/UV) +0.433 \times (Log(FIR/UV))^2
\end{equation}
where $FIR=1.26 \times (2.58 \times 10^{12} \times F_{60} + 10^{12} \times F_{100}) \times 10^{-26}     ~~~~~~[Wm^{-2}]$
\\$F_{60}$ and $F_{100}$ are the IRAS FIR fluxes (in Jy)
\\and $UV=10^{-3} \times 2000*10^{(UV_{mag}+21.175)/-2.5}	       ~~~~~[Wm^{-2}]$
\\(B4) can be obtained from a polynomial fitting inverting (B6): 
\begin{equation}
\tau_{UV}=(1/sec(i)) \times (0.0259+1.2002 \times A_{UV}+1.5543 \times A_{UV}^2-0.7409 \times A_{UV}^3 +0.2246 \times A_{UV}^4)
\end{equation}
Thus $A_\lambda$ are finally obtained. FIR/UV is available for 85 objects.
If FIR or UV measurements are unavailable we assume  $A_{UV}$ = 1.28; 0.85; 0.68 mag
for Sa-Sbc; Sc-Scd; Sd-Im-BCD galaxies respectively, as determined for galaxies 
in these three 
classes of Hubble type when FIR and UV measurements are available.\\
After the SEDs were corrected with the above method, we checked empirically that 
they do not contain a residual
dependence on the galaxy inclination. The corrected SEDs of 32 Sc galaxies,
binned in 4 intervals of inclination, and their fit parameters were found very consistent
one another.

\section {Fit of B\&C models to the data}

\begin{figure}[!t]
\epsscale{0.85}
\plotone{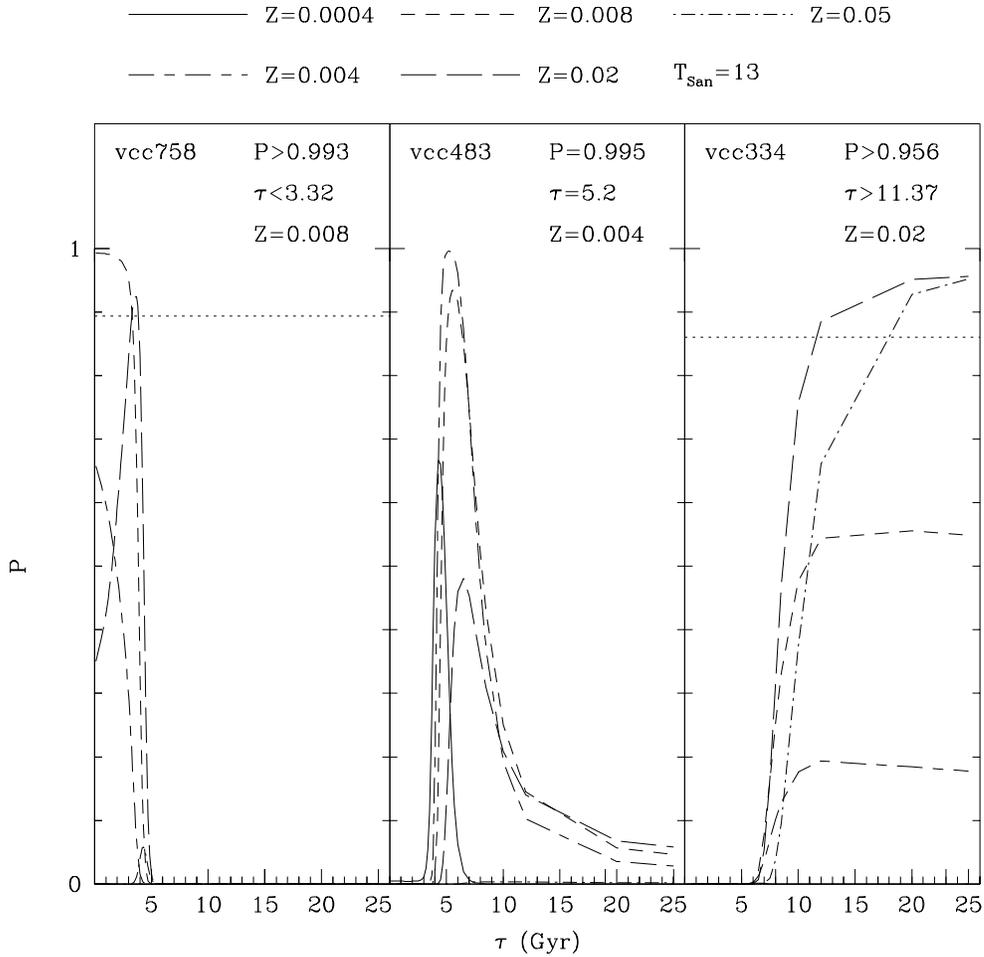}
\small{\caption{The fit probability as a function of $\tau$ for the various metallicities. An example of
P($\tau$) reaching a maximum for $\tau \to 0$ is given by VCC 758 (left). An example of
P($\tau$) reaching a maximum within the parameter space is given by VCC 66 (center). In this case
$\tau,Z_{max}$ are univocally determined from  $P_{max}(\tau,Z$). An example of
P($\tau$) reaching a maximum for $\tau \to \infty$ is given by VCC 159 (right). In the first and third case
$\tau,Z_{max}$ are determined using the thresholded probability of 0.90 of their maximum.}\label{prob}}
\end{figure}

The logarithmic fit of B\&C models to the data is obtained  minimizing 
the $\overline\chi^2$ (reduced chi-squared): 
\begin{equation}
\overline\chi^2=\frac{1}{d}\sum\limits_{i=1}^{N}\left(\frac{F_{i,obs}-F_{i,mod}}{\sigma_i}\right)^2
\end{equation}
where $d$ is the number of degrees of freedom, $d=n-v$, $n$ represents the 
number of bands for which $F_{i,obs}\neq0$, 
$v$ is the number of constraints, i.e. the number of fitted parameters
(e.g. $Z$ and $\tau$).
$F_{i,obs}$ and $F_{i,mod}$ are the observed and model fluxes respectively, and
$\sigma_i$ is the error associated with each photometric "band" (UV, U, B V, J, H, K, H$\alpha$, 
spectrum).
\begin{equation}
log F_{i,obs}= -0.4 \cdot Mag_i + C_i - A(f_i)~({\rm erg~cm^{-2}~sec^{-1}~\AA^{-1}})
\end{equation}
where\\
$C_i=-2.5\cdot \left(log(\int F_{Vega}(\lambda) R_i(\lambda) d\lambda\right)$\\
are tabulated in Table \ref{fconv_flux} and\\
$A(f_i)=-2.5\cdot log(\int R_i(\lambda) d\lambda)$\\
represent the area of the filter transmission (taken from B\&C).\\
For the  B, V, H bands we assume an uncertainty of 0.15 mag, 0.2 for
U, J and K, 0.4 mag for UV
and a large uncertainty of 1 mag for the $<912$ \AA~ flux derived from H$\alpha$.\\
The spectra were divided in 6 bins: 3800-4000; 4000-4200; 4200-5000; 5000-5800;
5800-6600; 6600-7000 \AA.
The first 2 narrow bins bracket the 4000 \AA~ break.
Each spectral bin contributes to the fit
as one "photometric point". The error on each point was determined by combining 
the rms measured in line-free regions of the individual spectral bins   
with 0.10 mag uncertainty in the photometric calibration.\\
From $\overline\chi^2$ we compute a fit probability defined as:
\begin{equation}
P(\overline\chi^2\geq\overline\chi^2_{obs})=\frac{2}{2^{d/2}\Gamma(d/2)}
\int^\infty_{\overline\chi^2_{obs}}x^{d-1}e^{-x^2/2}dx.
\end{equation}
where $\Gamma$ is the incomplete-gamma function.
$ P(\overline\chi^2\geq\overline\chi^2_{obs})$ is the probability of obtaining 
a $\overline\chi^2$ equal to the observed value 
($\overline\chi^2_{obs}$). Only fits with probability $>5\%$ are considered.
Only 3 galaxies in the photometric sample and none in the spectro-photometric sample
have $P<5\%$, while 85 \% of the objects have $P>70\%$.
For any given metallicity the probability of the fit is traced as a function of $\tau$.
For most cases the distribution reaches a maximum at a given $\tau$ and $Z$
which are then selected as best parameters (see Fig.\ref{prob} center).
In some cases the probability never reaches a maximum within the parameter space, but
tends asymptotically to a maximum either for $\tau \to 0$ or $\tau \to \infty$  
(see Fig.\ref{prob} left and right).
Because neither of these values is meaningful in these cases we adopt
$\tau$, $Z$ corresponding to a threshold probability of 0.90 of the maximum.\\
Two notes of caution.\\
1) In spite of our effort for homogenizing the apertures at which data
used in the present work were taken, 
total UV magnitudes, integrated spectra and broad band photometry  
have still slight aperture differences (see Section 3). The residual aperture effects 
on the SED shapes should however be within the quoted
photometric uncertainties, thus not affecting significantly the determination     
of the fit parameters.\\
2) The incompleteness in the various observations bands  
increases toward lower luminosities (see Section 2). This might bias the 
luminosity dependence of the fit parameters. 
We have checked this effect by excluding one by one
from the fit the UV, the U and B data and the spectra. We conclude 
that none of these exclusions produces significant changes. In particular excluding
the U, B and spectral data does nothing but adding some noise in the correlations. 
The exclusion of UV data has an appreciable effect on the SEDs of some early-type spirals 
which become "bluer", thus with longer $\tau$.

\begin{table}[!t]
\centerline{
\begin{tabular} {cc}
\hline
\noalign{\smallskip}
Filter& C \\
\noalign{\smallskip}
\hline 
\noalign{\smallskip}
 UV &   15.44\\
  U &   14.11\\
  B &   13.06\\
  V &   13.77\\
  J &   15.85\\
  H &   16.11\\
  K &   16.62\\
\noalign{\smallskip}
\hline
\end{tabular}}
\caption{\footnotesize  Conversion coefficients from magnitude to flux for
the considered filters. The coefficient for the UV filter corresponds to 2000 \AA~since
$\rm UV_{1650}$ magnitudes are converted to 2000 \AA using an average correction of 0.2 mag (see Section 
3.2).}\label{fconv_flux}
\end{table}   

\end{document}